\documentclass[11pt, letterpaper]{article}

\usepackage{graphicx}
\usepackage{amsmath}
\usepackage{amssymb}
\usepackage{amsmath}
\usepackage{amsfonts}
\usepackage{mathtools}
\usepackage[super]{nth}
\usepackage{setspace}
\usepackage{enumitem} 
\usepackage{epstopdf}
\usepackage{color}
\usepackage[letterpaper,left=1in,right=1in,top=1in,bottom=1in]{geometry}
\usepackage[]{algorithm2e}
\usepackage{multirow}
\usepackage{longtable}
\usepackage{rotating}
\usepackage{mathrsfs}
\usepackage{subfig}
\usepackage{float}
\usepackage{graphicx}
\usepackage{caption}

\graphicspath{{Figures//}}

\title{\Large \bf Impact of sensor placement in soil water estimation: A real-case study}

\author{
\centerline{\normalsize Erfan Orouskhani$^{a}$, Soumya R. Sahoo$^{a}$, Bernard T. Agyeman$^{a}$, Song Bo$^{a}$, Jinfeng Liu$^{a,}$\thanks{Corresponding author: J. Liu. Tel: +1-780-492-1317. Fax: +1-780-492-2881. Email: jinfeng@ualberta.ca.}}
\vspace{5mm}\\
\centerline{\small $^{a}$Department of Chemical \& Materials Engineering, University of Alberta,}\\
\centerline{\small Edmonton, AB T6G 1H9, Canada.}}
\allowdisplaybreaks

\begin{document}

\date{}
\maketitle

\setstretch{1.5}

\begin{abstract}
One of the essential elements in implementing a closed-loop irrigation system is soil moisture estimation based on a limited number of available sensors. One associated problem is the determination of the optimal locations to install the sensors such that good soil moisture estimation can be obtained. In our previous work, the modal degree of observability was employed to address the problem of optimal sensor placement for soil moisture estimation of agro-hydrological systems. It was demonstrated that the optimally placed sensors can improve the soil moisture estimation performance. However, it is unclear whether the optimal sensor placement can significantly improve the soil moisture estimation performance in actual applications. In this work, we investigate the impact of sensor placement in soil moisture estimation for an actual agricultural field in Lethbridge, Alberta, Canada. In an experiment on the studied field, 42 soil moisture sensors were installed at different depths to collect the soil moisture measurements for one growing season. A three-dimensional agro-hydrological model with heterogeneous soil parameters of the studied field is developed. The modal degree of observability is applied to the three-dimensional system to determine the optimal sensor locations. The extended Kalman filter (EKF) is chosen as the data assimilation tool to estimate the soil moisture content of the studied field. Soil moisture estimation results for different scenarios are obtained and analyzed to investigate the effects of sensor placement on the performance of soil moisture estimation in the actual applications.
\end{abstract}

\clearpage

\section{Introduction}

Freshwater scarcity is becoming a serious issue worldwide primarily due to population growth, climate change, and increasing pollution \cite{WWDRUN2009}. Of the total amount of freshwater, about 70\% is consumed in the agricultural activities, with the main consumer being irrigation \cite{UN_report}. Currently, the water-use efficiency in irrigation is estimated to be 60\% due to poor irrigation strategies \cite{WWDRUN2009}. In order to mitigate the freshwater supply crisis, the water-use efficiency in agriculture irrigation needs to be improved.
In the current irrigation practice, irrigation in general is determined in an open-loop fashion in which little real-time feedback from the field such as soil moisture is considered. The amount and time of irrigation are typically determined by the farmer based on their experience, which often leads to excessive or insufficient irrigation \cite{romero2012research}. 

One promising solution to address this issue and improve the water-use efficiency is to use a closed-loop irrigation system where a controller uses real-time field conditions to make the best irrigation decisions \cite{nahar2019closed}. Although using the closed-loop irrigation system can lead to optimized irrigation and increased crop yield and profit, its implementation can be challenging. In implementing a closed-loop irrigation system, the soil moisture information of the entire field which should be fed back to the controller is required. On the other hand, the agriculture fields usually are of very large scale and installing the sensors in the whole field is impractical. Therefore, one of the main barrier in implementing the closed-loop irrigation system is the lack of field-wide soil moisture measurements \cite{nahar2017observability}.

To address this issue, using state estimation techniques that reconstruct full states information based on the measurements of a small number of sensors have been proposed. Due to the nonlinearity of the field model, the nonlinear state estimators such as extended Kalman filter \cite{reichle2002extended,agyeman2020soil}, ensemble Kalman filter \cite{zhang2018comparison}, and particle filter \cite{pasetto2012ensemble,montzka2011hydraulic} have been typically used to address the problem of soil moisture estimation in the literature.
Reichle et al. \cite{reichle2002extended} compared the performance of the EKF and EnKF methods in the soil moisture estimation of a land surface model, known as the Catchment Model. Based on their study, although the performance of both filters was satisfactory, they had some drawbacks. EKF was computationally expensive due to the Jacobian matrix calculation during linearization, and EnKF required a large number of ensembles for good estimation performance. Walker et al. \cite{walker2001one} estimated the soil moisture of the simplified soil moisture model which was the linear version of the Darcy–Buckingham equation, by employing the linear Kalman filter and using real soil moisture measurements. Not considering the root water uptake term, using the simplified soil moisture model instead of Richards' equation, and investigating the soil moisture estimation for the only one-dimensional system were the shortcomings in their work. De Lannoy et al. \cite{de2007state} used EnKF with real field data to performed bias and regular soil moisture estimation. Although their approach could improve the overall performance of estimation, bias estimation in layers for which no observations were available was impossible.

In the above studies, the optimal sensor placement has not been considered. Because of the limited number of available sensors in the agricultural fields, it is an important problem to find the optimal location of the sensors in the soil such that improved state estimation can be obtained. 
In \cite{nahar2017observability}, Nahar and co-authors proposed to use the observability analysis to find the optimal sensor locations. However, the applicability of this method was restricted to one-dimensional systems. Then, in our recent work \cite{sahoo2019optimal}, the optimal sensor placement problem has been addressed by employing the modal degree of observability. 
It was found that optimally placed sensors can lead to much-improved soil moisture estimation performance. However, it is unclear whether the significantly improved estimation performance can still be observed in the actual applications. 
Moreover, in all the above studies, homogeneous soil parameters or simple arrangements of different soil types have been considered.

In this work, we consider an actual agricultural field in Lethbridge, Alberta, Canada, and investigate the impact of sensor placement in soil water estimation performance of the actual field. Soil moisture measurements from 42 soil moisture sensors installed at different depths were collected for one growing season. First, a three-dimensional agro-hydrological model with heterogeneous soil parameters of the studied field is developed. Then, a state estimator designed based on the extended Kalman filter (EKF) is adopted to estimate the soil water content. Subsequently, we apply the modal degree of observability to the three-dimensional system and determine where the best sensor locations are. Different scenarios are considered to estimate the soil water content of the studied field and the estimation results are analyzed to investigate the effect of sensor placement on the performance of soil moisture estimation in actual applications.


Some preliminary results of this paper were reported in \cite{ErfanConference2022}. 
Compared with \cite{ErfanConference2022}, this paper presents significantly detailed explanations and more simulation results for different scenarios of placing sensors in order to extensively evaluate the effect of sensor placement on the estimation performance in the actual application.

\section{Description of the studied field}
\label{Section 2}

\begin{figure}[h!]
	\centering
	\subfloat[The center pivot irrigation system of the field]{
		\includegraphics[width=0.45\textwidth]{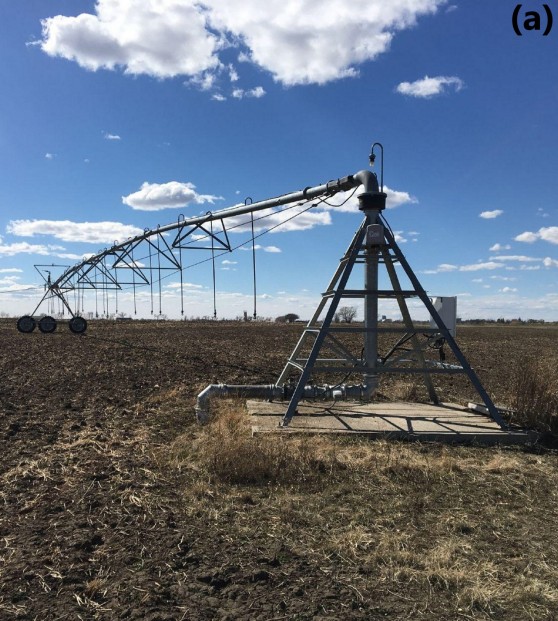}
		\label{fig:demo1}
	} %
	\qquad
	\subfloat[Location of the sensors in the studied field]{
		\includegraphics[width=0.45\textwidth]{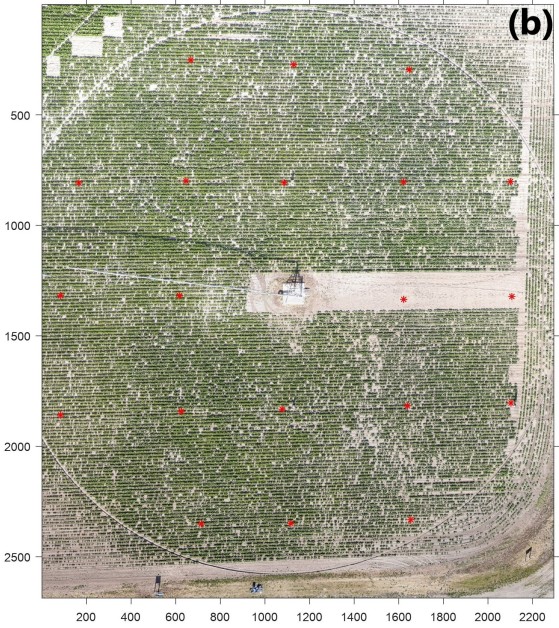}
		\label{fig:demo2}
	}
	\caption{Studied field in Lethbridge}
\end{figure}

The agricultural field studied in this work is located in Lethbridge, Alberta, Canada (Lon: -112.7385 : -112.7365, Lat: 49.6896 : 49.6908). The field is a circular one with a radius of about 50 meters. The depth of the field is 75 cm in the simulations of this work. One weather station is located near the agricultural studied field managed by Lethbridge Demo Farm Irrigation Management Climate Information Network (IMCIN). The weather station's data including the precipitation, wind speed, and air temperature can be obtained from the Alberta Climate Information Service (ACIS) website (https://agriculture.alberta.ca/acis/). The soil texture consists of three types of soil: clay, silt, and sand. Each area of the field has a different percentage of the soil types that makes the soil profile heterogeneous. Thus, the soil profile of the field has different properties at various zones. For example, on the left side of the field where the percentage of the clay in the soil is higher than other areas of the field, the water infiltration to the root zone is slower compared to other areas. In the studied field, a centre pivot is used as the irrigation implementing system as shown in Figure 1(a). In irrigation time, the center pivot rotates at a speed of 0.011 m/s. 

The soil profiles at 60 points of the studied field (20 points from surface to depth 25 cm, 20 points at 25 cm to 50 cm, and 20 points at 50 cm to 75 cm) were sampled. After collecting the soil sample, the soil properties of the sampling points including the wilting point, the electrical conductivity of the water, and the percentage of the clay, silt, and sand existing in the soil samples were estimated in the soil lab. The soil profile data will be used in Section \ref{subsection parameters} to interpolate the soil parameters of the entire field. 

\begin{figure}
	\centerline{\includegraphics[width=0.96\textwidth]{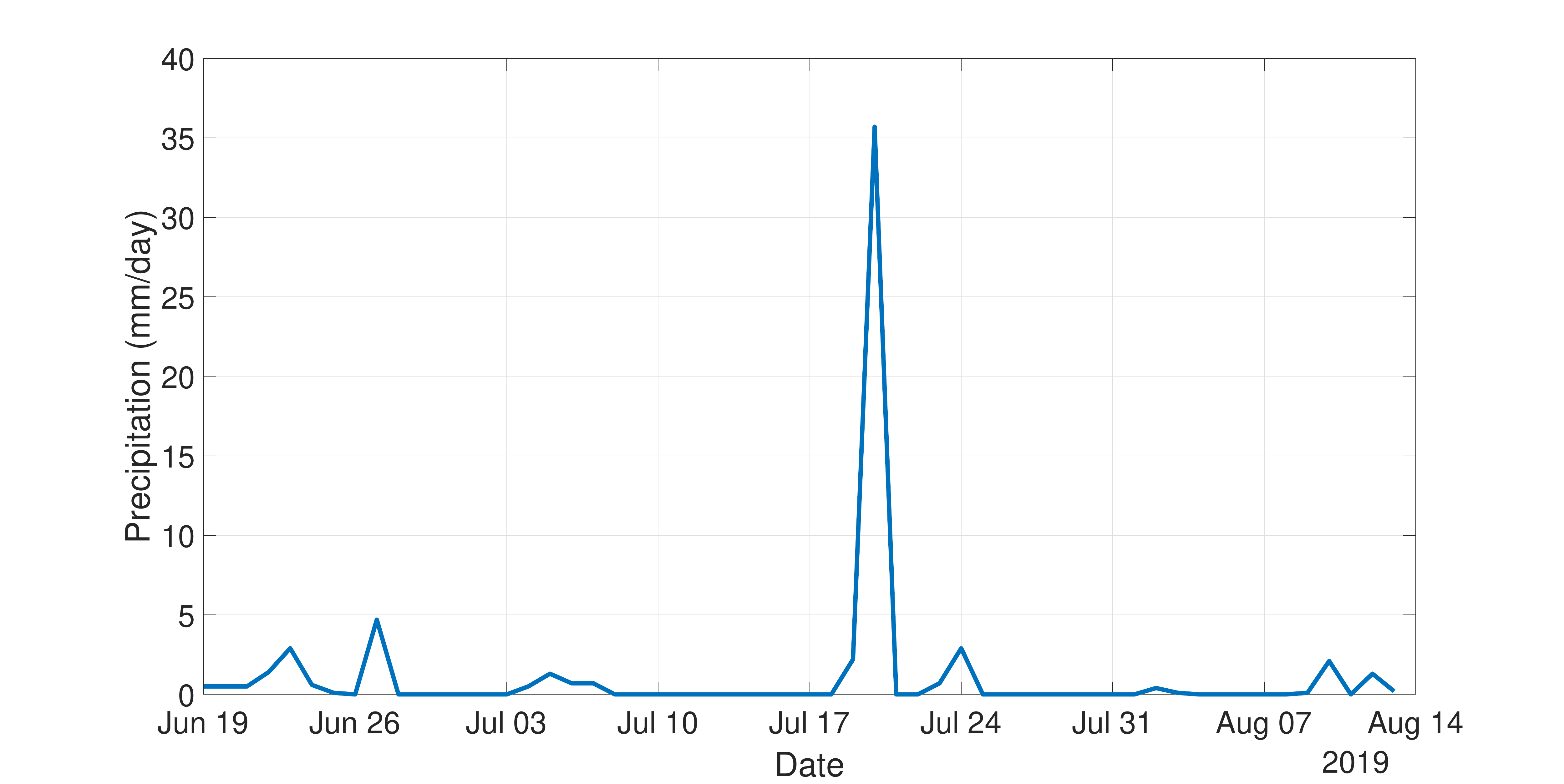}}
	\caption{Daily precipitation data of the studied field during the period under investigation}
	\label{fig:rain}
\end{figure}

Moreover, 42 watermark sensors were installed in the field at different depths (14 sensors at the depth of 25 cm, 14 sensors at the depth of 50 cm, and 14 sensors at the depth of 75 cm, below the surface) to measure the soil water tension of these locations. The measurements were collected every 30 minutes from June 19 to August 13, 2019. Figure 1(b) shows the location of the watermark sensors in the studied field. 
During the experiment, a data logger was used to collect the data from the watermark sensor. Since the data logger did not have enough ports to connect all forty-two sensors, a multiplexer was used to connect the sensors to a single data logger. In the whole field, we employed two data loggers and two multiplexers for data collection. Two solar panels were also installed to charge the data loggers. Some irregular features in the collected data set were observed, which increase the model plant mismatch and cause the overall soil moisture estimation more challenging. The collected data will be used in Section 7 to estimate the soil moisture of the entire field through a state estimator. The precipitation data of the studied field obtained from ACIS is illustrated in Figure \ref{fig:rain} on a daily basis for the period under investigation. 

\section{Modeling of the water dynamics of the studied field}
\label{Section 3}

\subsection{Agro-hydrological system description}

\begin{figure}[h!]
	\centerline{\includegraphics[width=0.7\textwidth]{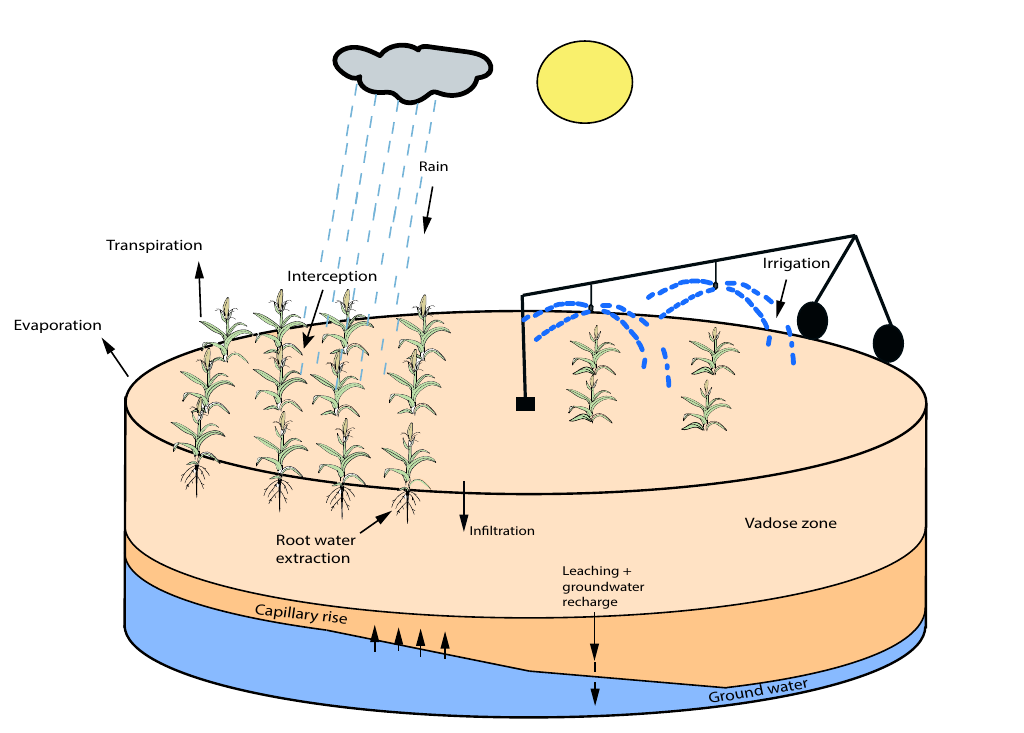}}
	\caption{An agro-hydrological system \cite{agyeman2020soil}}
	\label{fig:agro}
\end{figure}

An agro-hydrological model characterizes the hydrological cycle between the soil, the water, the atmosphere, and the crop. In this work, the three-dimensional agro-hydrological model is considered in which the water inflows to the system are rainfall and irrigation, and the system outflows are evapotranspiration, runoff, and drainage \cite{nahar2019closed,bo2020decentralized}. Figure \ref{fig:agro} provides an illustration of an agro-hydrological system \cite{agyeman2020soil}.

The dynamics of soil water can be modeled using the Richards' equation as follows \cite{richards1931capillary}:
\begin{eqnarray}
	\frac{\partial \theta}{\partial t}=C(h)\frac{\partial h}{\partial t}=\nabla\cdot(K(h)\nabla(h+z)) - S
	\label{eq:Richards}
\end{eqnarray}
where $h ~(m)$ is the pressure head, $\theta ~(m^3 m^{-3})$ is the volumetric water content, $t~(s)$ is time, $z~(m)$ is the spatial coordinate, $K(h)~(ms^{-1})$ is the unsaturated hydraulic water conductivity, $ C(h)~(m^{-1})$ is the capillary capacity, and $S~(m^3 m^{-3}s^{-1})$ denotes the sink term, representing the root water extraction rate. 
In equation (\ref{eq:Richards}), the soil hydraulic functions $\theta (h)$, $K(h)$, and $C(h)$ can be obtained by the Mualem-van Genucthen model \cite{VanGenuchten1980}:
\begin{eqnarray}
	\theta (h)=\theta_r +(\theta_s-\theta_r)\bigg[\frac{1}{1+(-\alpha h)^n}\bigg]^{1-\frac{1}{n}}
\end{eqnarray}
\begin{eqnarray}
K(h)=K_s\bigg[(1+(-\alpha h)^n)^{-\big(\frac{n-1}{n}\big)}\bigg]^{\frac{1}{2}}\times\Bigg[1-\bigg[1-\Big[(1+(-\alpha h)^n)^{-\big(\frac{n-1}{n}\big)}\Big]^{\frac{n}{n-1}}\bigg]^{\frac{n-1}{n}}\Bigg]^2
\end{eqnarray}
\begin{eqnarray}
	C(h)=(\theta_s-\theta_r)~\alpha n~\bigg(1-\frac{1}{n}\bigg)~(-\alpha h)^{n-1}\big[1+(-\alpha h)^n\big]^{-\big(2-\frac{1}{n}\big)}
\end{eqnarray}
where $\theta_s~(m^3m^{-3})$, $\theta_r~(m^3m^{-3})$, $K_s~(ms^{-1})$ are the saturated volumetric moisture content, residual moisture content and saturated hydraulic conductivity, respectively. $n$ and $\alpha$ are curve-fitting soil hyrdraulic properties. 
The parameters $\theta_s, \theta_r, K_s, \alpha$, and $n$ form a set of soil hydraulic parameters that determine the soil properties of the field.

\subsection{Interpolation of soil parameters}
\label{subsection parameters}

\begin{figure}[h!]
	
	\centering
	\subfloat[$\theta_s~(m^3m^{-3})$]{
		\includegraphics[width=0.48\textwidth]{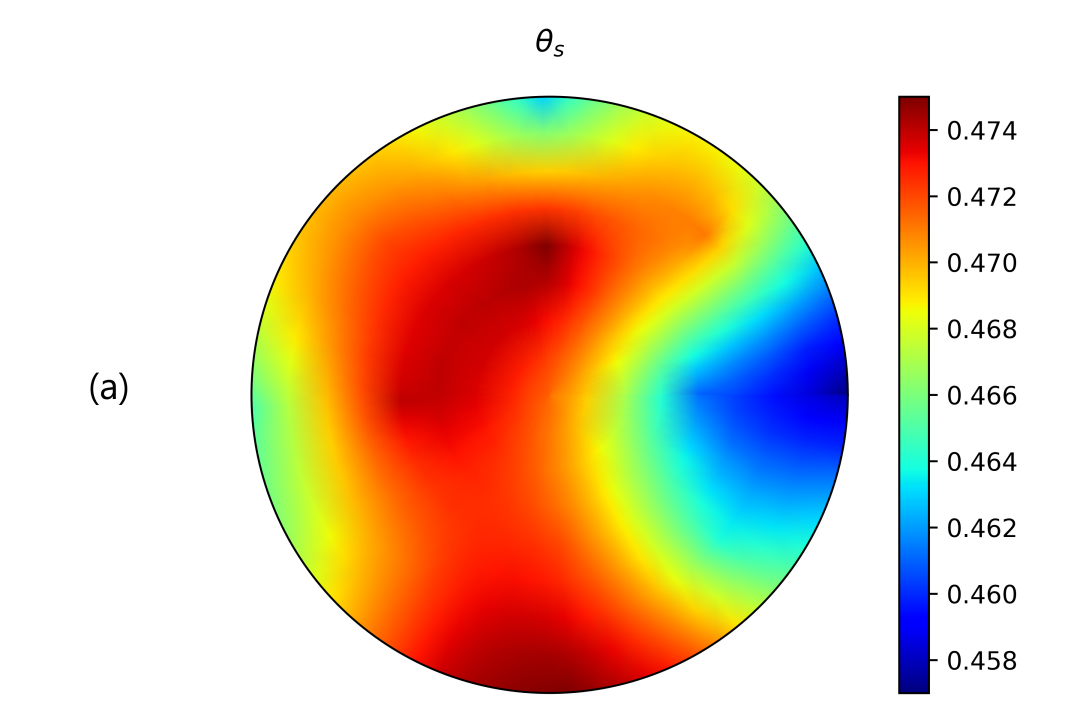}
	} 
	\subfloat[$\theta_r~(m^3m^{-3})$]{
		\includegraphics[width=0.48\textwidth]{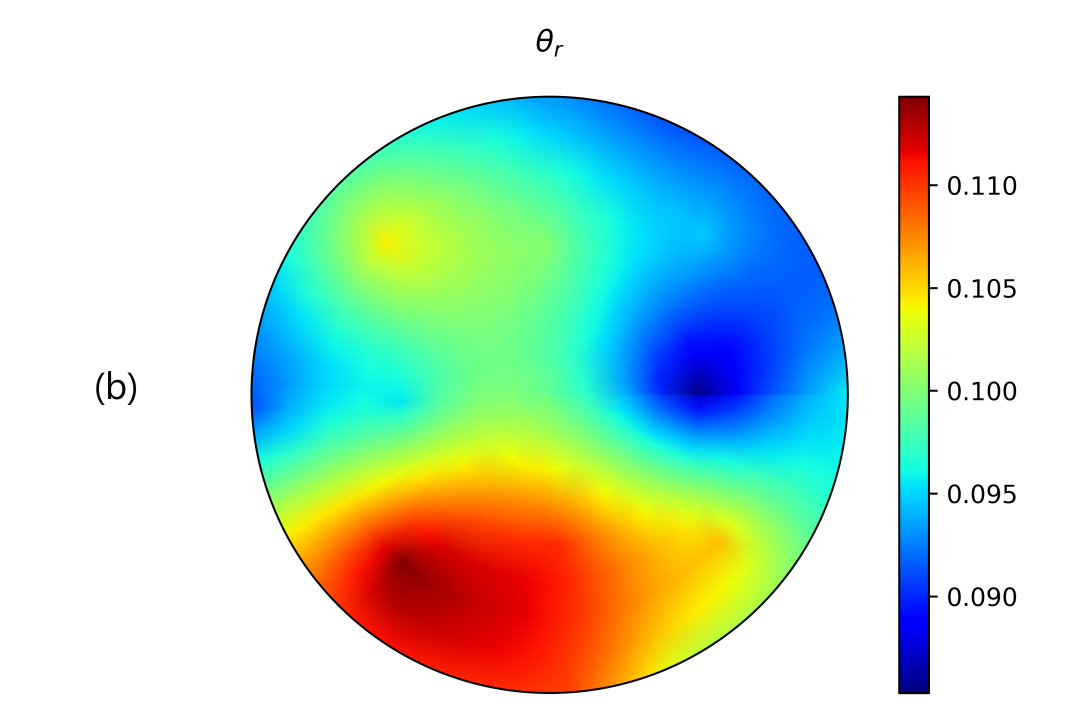}
	}
	
	\centering
	\subfloat[$K_s~(cmhr^{-1})$]{
		\includegraphics[width=0.48\textwidth]{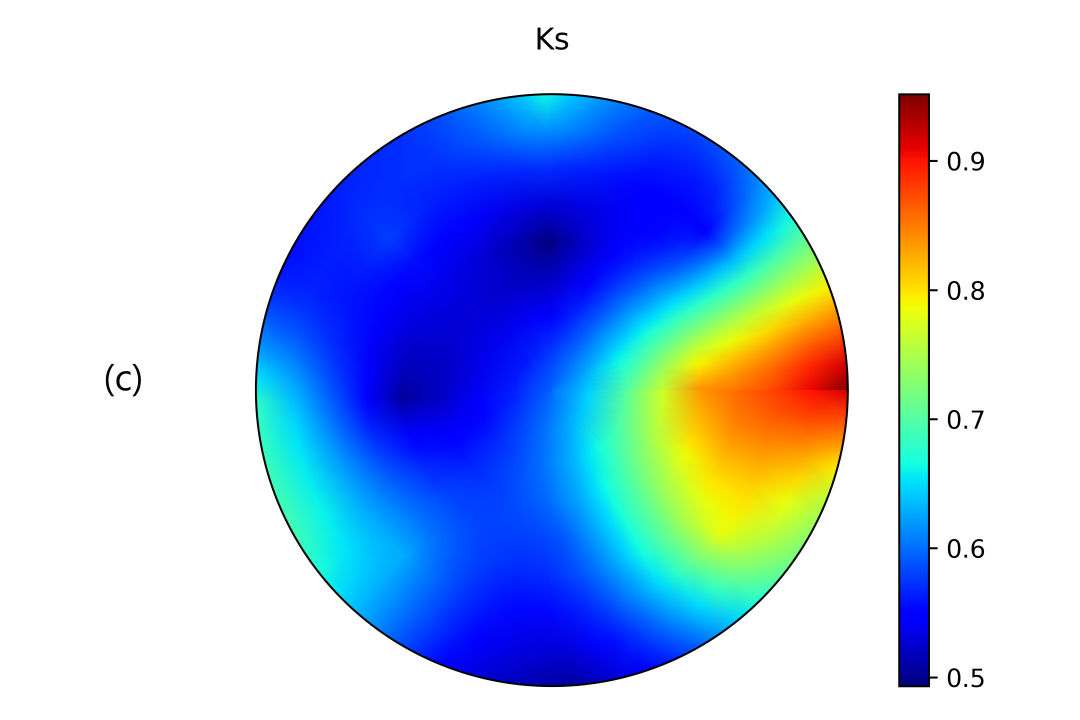}
	}
	\subfloat[$\alpha~(m^{-1})$]{
		\includegraphics[width=0.48\textwidth]{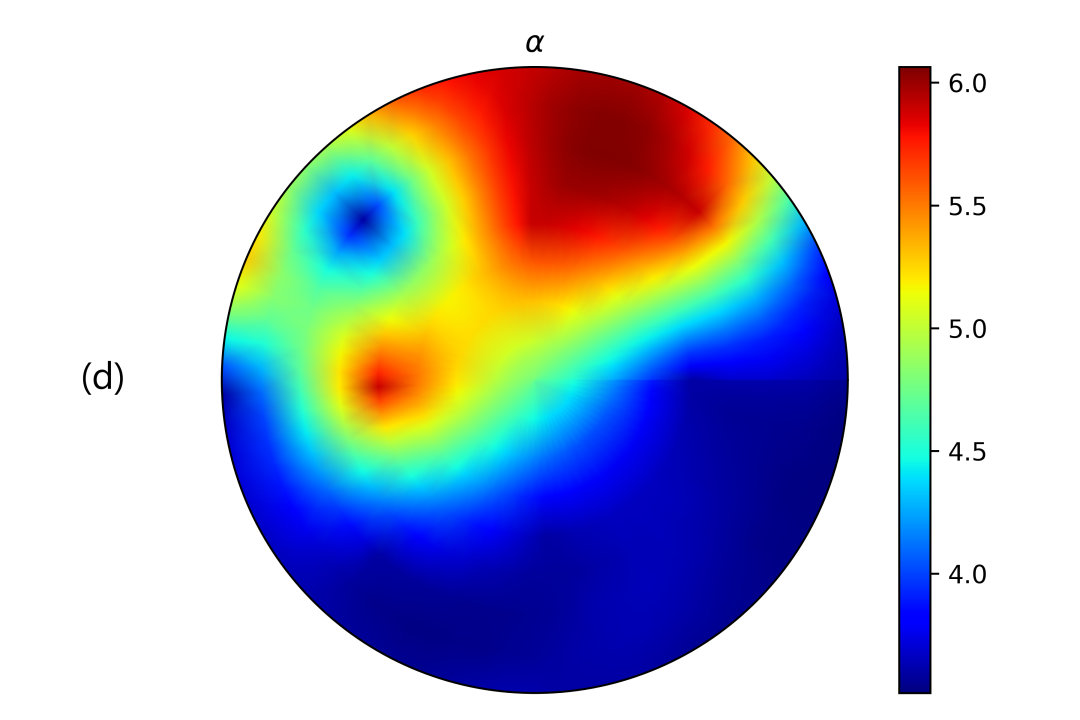}
	} 
	
	\centering
	\subfloat[$n$]{
		\includegraphics[width=0.48\textwidth]{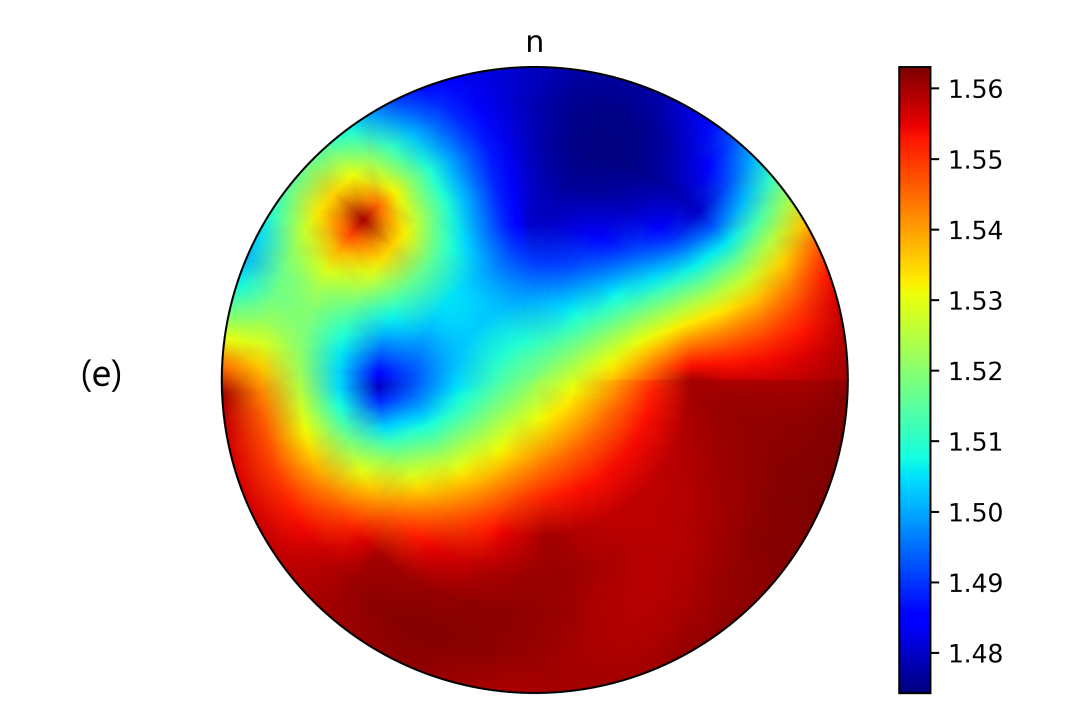}
	}

	\caption{Heterogeneous distribution of soil parameters on the surface of the studied field obtained from the Kriging interpolation}
	\label{fig:soil_para1}
	
\end{figure}

Saturated hydraulic conductivity $K_{s}~(\frac{m}{s})$, saturated soil moisture $\theta_{s}~(\frac{m^{3}}{m^{3}})$, residual soil moisture $\theta_{r}~(\frac{m^{3}}{m^{3}})$, and curve-fitting soil hydraulic properties $\alpha~(\frac{1}{m})$ and $n$ are the soil parameters of the model. 
Each type of the soil has its own set of soil parameters. 
Due to the heterogeneity of the soil in the studied field, the soil parameters are different at different points of the field. In fact, each point in the field which corresponds to a node in the discretized model has its own set of soil parameters.   
These soil parameters are unknown and need to be obtained.
In this work, we use the Kriging interpolation method \cite{matheron1963principles} to estimate the soil parameters of the entire field.
We first used the 60 soil samples of the studied field and determined the soil texture type of the sampled points by measuring the percentage of the clay, silt, and sand soils existing in the samples. Next, we obtained the set of soil parameters for these sampling points based on the composition of the soil types \cite{carsel1988developing}. Subsequently, we used the soil parameters of these 60 sampled points as the measurements in the Kriging interpolation method to interpolate the soil parameters of the entire field.  Figure~\ref{fig:soil_para1} shows the interpolated soil parameters of the surface of the studied field. The results show that the soil parameters of the field are heterogeneous. 

\subsection{Polar form of Richards' equation}

In \cite{agyeman2020soil}, it was demonstrated that the cylindrical coordinate version of the Richards' equation is very suitable for the modeling of an agricultural field equipped with a center pivot irrigation system due to its ability to account for the circular movement of the center pivot. 
Therefore, since the center pivot is used as the irrigation implementing system in the studied field, we will use the cylindrical coordinate of the Richards' equation in this work.     
Cylindrical coordinate representation of the Richards’ equation is expressed as follows \cite{agyeman2020soil}:
\begin{eqnarray}
C(h)\frac{\partial h}{\partial t}=\frac{1}{r}\frac{\partial}{\partial r}\bigg[r K(h)\frac{\partial h}{\partial r}\bigg]+\frac{1}{r}\frac{\partial}{\partial \theta}\bigg[\frac{K(h)}{r}\frac{\partial h}{\partial \theta}\bigg]+
\frac{\partial}{\partial z}\bigg[K(h)\bigg(\frac{\partial h  }{\partial z}+ 1\bigg)\bigg]-S
\label{eq:RichardsPolar}
\end{eqnarray}
where  $r,\theta,z$ represent the radial, azimuthal, and axial directions, respectively. 
Eq. (\ref{eq:RichardsPolar}) is a nonlinear parabolic-elliptical partial differential equation (PDE) with respect to the temporal ($t$) and  the spatial variables ($r,~\theta,~z$).  

\subsection{Model discretization}
\label{subsection Model}
Obtaining an analytical solution to Eq. (\ref{eq:RichardsPolar}) is difficult due to its nonlinearity, thus numerical solutions are needed to solve this equation. In \cite{agyeman2020soil}, firstly the two point central finite difference scheme was employed to approximate the derivatives of Eq. (\ref{eq:RichardsPolar}) with respect to the spatial variables {\footnotesize( $r,~\theta,~z$)}. This converts the PDE into a set of ordinary differential equation (ODE) in terms of the temporal variable ($t$). Then, the Backward Differentiation Formulas (BDFs) methods were used to approximate the time derivative in the resulting ODEs. To implement the BDFs methods, the `cvodes' integrator in CasAdi (version 3.5.1) was used \cite{agyeman2020soil}. 
Also, some boundary conditions were imposed to the system to solve Eq. (\ref{eq:RichardsPolar}) numerically. For example, the symmetry boundary condition {\footnotesize $\frac{\partial h}{\partial r}|_C=0$} is used at the centre ({\footnotesize$C$}) of the field to deal with singularity that occurs at {\footnotesize$r=0$}; the Neuman boundary condition {\footnotesize$\frac{\partial (h)}{\partial z}|_T=-1-\frac{U_{irr}}{K(h)}$} is used at the top ({\footnotesize $T$}) of the field at {\footnotesize$z=0$} to incorporate the irrigation rate {\footnotesize$U_{irr} (ms^{-1})$} into the Richards' equation.  
The same numerical model development and discretization scheme is used in this paper. Specifically, we discretize the field into 6, 40 and 22 nodes in the radial, azimuthal and axial directions, respectively. The head pressure of the soil at these discretized nodes are the states of the system.

\subsection{State-space representation of the field model}
The field model is expressed in state space form as:
\begin{eqnarray}
\label{eq:original system}
\dot{x}(t)=F(x(t),u(t)) + \omega(t)
\end{eqnarray}
where $x(t) \in \mathbb{R}^{N_{x}}$ represents the state vector containing $N_{x}=5,280$ pressure head values for the corresponding spatial nodes.  $u(t) \in \mathbb{R}^{N_{u}}$ and $\omega(t) \in \mathbb{R}^{N_{x}}$ represent the input vector and the model disturbances respectively. Specifically, in this work, the sensors directly measure the states of the system and the output vector $y(k)$ is the head pressure ($h$) at the measured nodes of the field. Thus, the output equation simply represents a matrix ($C$) indicating which states are measured by the sensors:
\begin{eqnarray}
y(t)= C x(t) + v(t) \label{eq:original system2}
\end{eqnarray}
where $y(t) \in \mathbb{R}^{N_{y}}$ and $v(t) \in \mathbb{R}^{N_{y}}$ respectively denote the measurement vector and the measurement noise.  The matrix $C$ is determined by the sensor placement algorithm.

\section{Optimal sensor placement}
\label{Section 4}

In order to determine the best locations to install the sensors in the agricultural fields, Sahoo and co-authors proposed to use the modal degree of observability \cite{sahoo2019optimal}. 
They demonstrated that the degree of observability tells us how strongly or weakly observable a system is and it can be used as a measure of the optimality of sensor placement. 
In this paper, we use the algorithm presented in \cite{sahoo2019optimal}. In the following, we summarize the algorithm.

Modal degree of observability inspired by the PBH test analyzes the ability of a sensor node to estimate other nodes of the system. In the PBH test, if the entry of the right eigenvector $v_{ij}$ is zero, then the $j^{th}$ node is not observable by measuring the $i^{th}$ node. Based on the extension of this test, Gu et al \cite{gu2015controllability} proposed that the node $j$ is weakly observable from the sensor node $i$, if the entry of $v_{ij}$ is small. Thus, this approach is able to find the nodes that are difficult to estimate from a sensor node and subsequently helps us to find the optimal sensor locations. For a node $i$ at a specific time $k$, the normalized measure of the modal degree of observability can be calculated as~\cite{sahoo2020optimal}:
\begin{eqnarray}
\label{eq:sensor_placement}
O_{i}^{(k)} = \sum_{j=1}^{n} (1- \lambda_{j}^{2} (A_{d}^{(k)})) v_{ij}^{2}
\end{eqnarray}
where $A_{d}^{(k)}$ is the discretized model Jacobian matrix at time $k$ that can be obtained from $A_{d}^{(k)} = e^{A(k)T}$ when $T$ is the sampling time, and $\lambda_j ( j = 1,…,n)$ are the eigenvalues of matrix $A_{d}^{(k)}$. 
Based on the definition \cite{sahoo2019optimal}, the degree of observability of the system is the highest when the sensors are located at nodes with the highest degree of observability. Thus, the determination of the optimal sensor placement which is based on the maximization of the degree of observability, consists of three steps:
\begin{enumerate}[label={(\arabic*)}]
	\item At a time instant $k$, calculate the normalized measure of the modal degree of observability $O_{i}^{(k)}$ for all the system nodes $i, i = 1,…,n$, where $n$ is the total number of the states.
	\item Compute the final modal degree of observability ($O_{i}$) for each node as the average value of the modal degree of observability over all the time instants.
	\item Order the measures $O_i, i = 1,…,n$, according to their values. The optimal locations to place the sensors are the nodes with the highest $O_i$ values.
\end{enumerate}

Since in this method it is not required to consider all the combinations of the sensors and in order to determine the optimal sensor placement we only have to calculate the $O_i$ values for all the states, order them, and find the biggest $O_i$ value, this approach is computationally very efficient especially for large-scale systems such as the three-dimensional agro-hydrological system.
In the following, we describe how the above sensor placement algorithm can be applied to the system considered in this work.

Firstly, the system in Eq. (\ref{eq:original system}) is simulated numerically and the state trajectory ($x(t)$) of the system is obtained during the simulation. Then, a symbolic approach using CasAdi is employed to calculate the Jacobian matrix ($A$) which is required in implementation of the optimal sensor placement algorithm. In calculation of Jacobian matrix, the state trajectory obtained from the previous step is required at each operating point ($k$). 
Next, we calculate the discretized model Jacobian matrix from $A_d(k) = e^{A(k)T}$ and use it in Eq. (\ref{eq:sensor_placement}) to obtain the degree of observability for all nodes of the system at a specific operating point. Eventually, the final modal degree of observability for each node is the average of the modal degree of observability values obtained at the operating points.

\begin{figure}
	\centering
	\centerline{\includegraphics[width=0.9\textwidth]{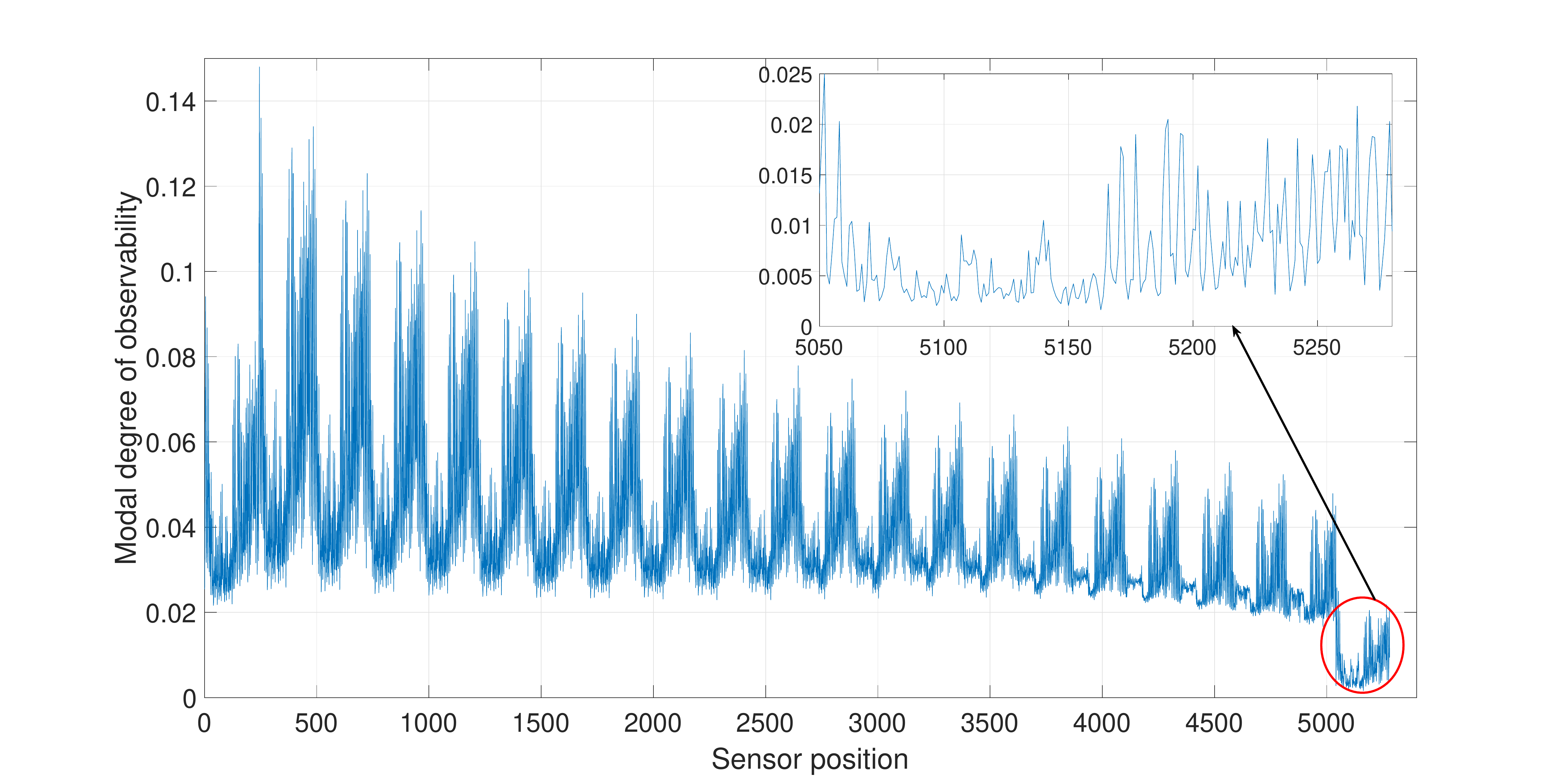}}
	\caption{Average modal degree of observability for different nodes of the system}
	\label{fig:degree1}
\end{figure}

Figure \ref{fig:degree1} represents the modal degree of observability for different nodes of the system considered in this work. From Figure \ref{fig:degree1}, it can be seen that nodes between 240 and 480, located at 65 cm below the surface layer, have relatively higher values of the modal degree of observability around 0.0574, while placing sensors on the surface corresponding to nodes 5040 to 5280 gives the lowest modal degree of observability about 0.0075.
Furthermore, the location of the optimal sensor placement is node 244 which has the highest degree of observability value, 0.148.


\section{Soil moisture estimator design}
\label{Section 5}

Once the system observability is checked and the optimal sensor placement is found, state estimation can be performed. The nonlinear state estimators should be used to estimate the states of the agro-hydrological model due to its nonlinearity.  
In this work, we choose the discrete-time extended Kalman filter (EKF) to estimate the states. 
The EKF algorithm consists of two steps, the prediction step, and the update step. In the prediction step, the state $x$ and its covariance matrix $P$ are predicted using the model of the system. In the update step, the prediction values $x$ and $P$ are updated using the actual measurements. 
The detail steps are described as follows:

\textbf{Initialisation}
\begin{enumerate}[label={(\arabic*)}]
	\item The continuous-time system, Eqs. (\ref{eq:original system}) and (\ref{eq:original system2}), is discretized to obtain its discrete-time equivalent. The discrete-time version can be expressed as:
	\begin{eqnarray*}
		x_{k+1} = f(x_{k},u_{k}) + \omega_{k}\\
		y_k = C x_{k} + v_{k}
	\end{eqnarray*}
	and the filter is initialized with $\hat{x}_{0}$ and $P_{0|0}$.
\end{enumerate}

\textbf{Prediction}
\begin{enumerate}[label={(\arabic*)}]
	\item The new state of the system is predicted at time $t_{k+1}$, using the previous state estimate $\hat{x}_{k|k}$ and its
    covariance matrix $P_{k|k}$, and the new input $u_k$ to the system: 
	\begin{gather*}
		\hat{x}_{k+1|k}=f(\hat{x}_{k|k},u_k)
	\end{gather*}
	\item The state covariance matrix is obtained by
	\begin{gather*}
		P_{k+1|k}=A_kP_{k|k}A_k^T+Q
	\end{gather*}
	where $A_k=\frac{\partial f}{\partial x}\big|_{\hat{x}_{k|k},~u_k}$ and Q is the covariance matrix of the process disturbance $\omega$. In this work, we use a symbolic approach using CasAdi to calculate the Jacobian matrix ($A$).
\end{enumerate}

\textbf{Filtering}
\begin{enumerate}[label={(\arabic*)}]
	\item 
	We use the observation $y_{k+1}$ at time $t_{k+1}$ to update the state and its covariance matrix. The observation vector includes the states are measured by the sensors. The location of the sensors is determined by the optimal sensor placement algorithm. Kalman gain matrix, $G_{k+1}$ can be calculated as
	\begin{eqnarray*}
		G_{k+1}=P_{k+1|k}C^T[C P_{k+1|k}C^T + R]^{-1}
	\end{eqnarray*}
	where $R$ is the covariance matrix of the measurement noise $v$.
	\item Once the updated Kalman gain is obtained, the state is updated:
	\begin{gather*}
		\hat{x}_{k+1|k+1}=\hat{x}_{k+1|k} + G_{k+1}[y_{k+1}-C \hat{x}_{k+1|k}]
	\end{gather*}
	\item The state covariance matrix is updated as follows
	\begin{gather*}
		P_{k+1|k+1}=[I-G_{k+1}C]P_{k+1|k}
	\end{gather*}
\end{enumerate}

In this work, we rely on extensive simulations to determine the appropriate tuning EKF parameters (matrices $P, Q$ and $R$). We examine the estimated state trajectories and estimation error for different tuning parameters and choose tuning matrices that improve significantly the estimation performance and result in a smaller estimation error. 
In the tuning matrix $P$, it is notable to mention that since our knowledge of the initial estimate of the state $\hat{x}_{0}$ is limited in the real case study, a high initial covariance matrix ($P_{0|0}=\infty I$) must be chosen.

\section{Initial simulation study}
\label{Section 6}

In this section, we evaluate the modal degree of observability results for large-scale three-dimensional agro-hydrological systems using state estimation with simulated data.
The three-dimensional agro-hydrological system obtained in Section \ref{Section 3}, is used to simulate the model and obtain the head pressure of the actual system and is further used in the prediction step of the estimator. In the agro-hydrological model, the interpolated soil parameters obtained in Section \ref{subsection parameters} are used as the parameters of the model. Thus, heterogeneous soil parameters are considered in the simulations. 
In addition to the soil parameters, the initial condition of the head pressure ($x_0$) is also non-uniform and each state of the system has a different initial condition. In this study case, the initial condition of the states in the actual system is a random variable between -0.95 m and -0.8 m.  
The irrigation amount is a constant rate of 3.6 mm/day which is applied to the farm in the first 8 hours of each day, between 0:00 AM to 8:00 AM.

\begin{figure}
	\centering
	\centerline{\includegraphics[width=0.66\textwidth]{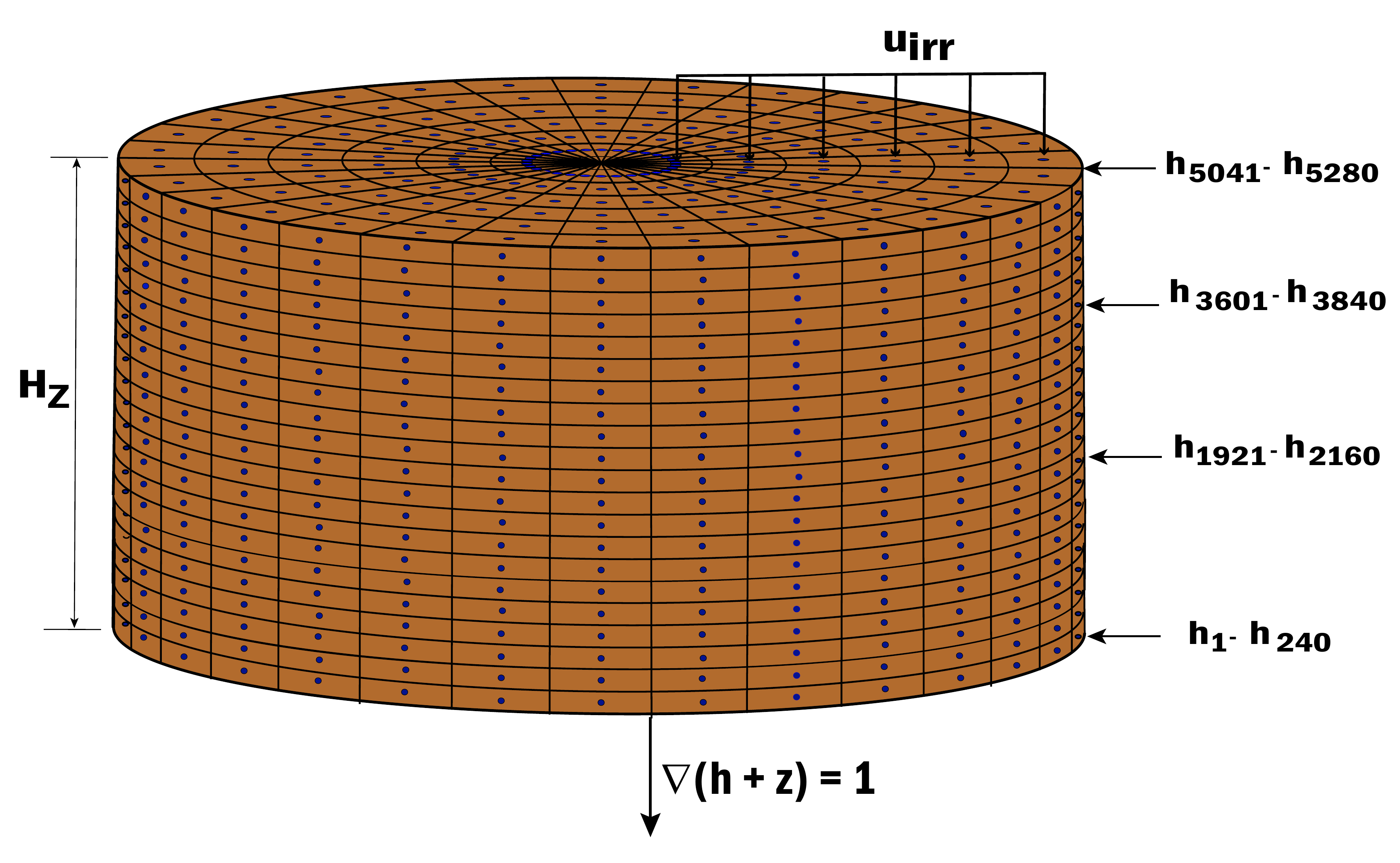}}
	\caption{A schematic diagram of the studied field \cite{agyeman2020soil}}
	\label{fig:Mesh}
\end{figure}

As we discussed in Section \ref{subsection Model}, the studied field is discretized into 5280 nodes (states) with 6 nodes in the radial direction, 40 nodes in the azimuthal direction, and 22 nodes in the axial direction. 
The first reason to choose these number of nodes for discretizing the field is to produce the nodes in the model that are matched with the actual measurement locations in the actual field.
In fact, the number of nodes in the radial and azimuthal directions and hence $\Delta r$ and $\Delta \theta$ were selected based on the location of the forty-two watermark sensors in the studied field. 
In addition, it was observed that further mesh refinement in any of the three directions did not result in a significant change in the state trajectories.   
Thus, it is considered that an accurate numerical approximation of Eq. \ref{eq:RichardsPolar} can be achieved with 5,280 states. 
Figure \ref{fig:Mesh} shows a schematic diagram of the studied field with its mesh structure, from the Reference \cite{agyeman2020soil}.
Additionally, based on the fact that the center pivot of the studied field takes about 8 hours to fully traverse the whole field and we divided the whole field in the azimuthal direction into 40 compartments, the appropriate time step size for the temporal discretization is about 12 minutes. 

In EKF design, 20\% mismatch in the initial condition of each state is considered. Twelve head pressure measurements at fixed locations in the field are used to correct the prediction state estimates in the update step of the EKF at each sampling time. Process noise and measurement noise are considered in the simulations and they have zero mean and standard deviations of $1 \times 10^{-6}$ and $6 \times 10^{-2}$, respectively. 
In the following simulations, we will compare the trajectories of the actual states and EKF estimated states for some selected nodes in order to observe the ability of the EKF for tracking the actual states and investigate the effect of sensor placement on the performance of state estimation.  
Additionally, the root mean square error (RMSE) at a time instant and the average RMSE will be calculated to assess the estimation performance:
\begin{equation}
RMSE_{x_{a}}(k) = \sqrt{\frac{\sum_{i=1}^{n_{x_{a}}}(\hat{x}_{a,i}(k) - x_{a,i}(k))^{2}}{n_{x_{a}}}}
\end{equation}
\begin{equation}
RMSE_{x_{a}} = \frac{\sum_{k=0}^{N_{sim}-1}RMSE_{x_{a}}(k)}{N_{sim}}
\end{equation}
where RMSE$_{x_{a}}(k)$ with $k = 0, \cdots, Nsim-1$ shows the evolution of the RMSE value over time and RMSE$_{x_{a}}$ shows the average value.
In order to fairly compare the performance of state estimation between different scenarios we use the normalized RMSE {\footnotesize ($NRMES = \frac{RMSE}{\bar{y}}$)} which facilitates the comparison between datasets or models with different scales.

\begin{figure}
	\centering
	\subfloat[State trajectory at depth = 5 cm]{ 
\includegraphics[width=0.46\textwidth]{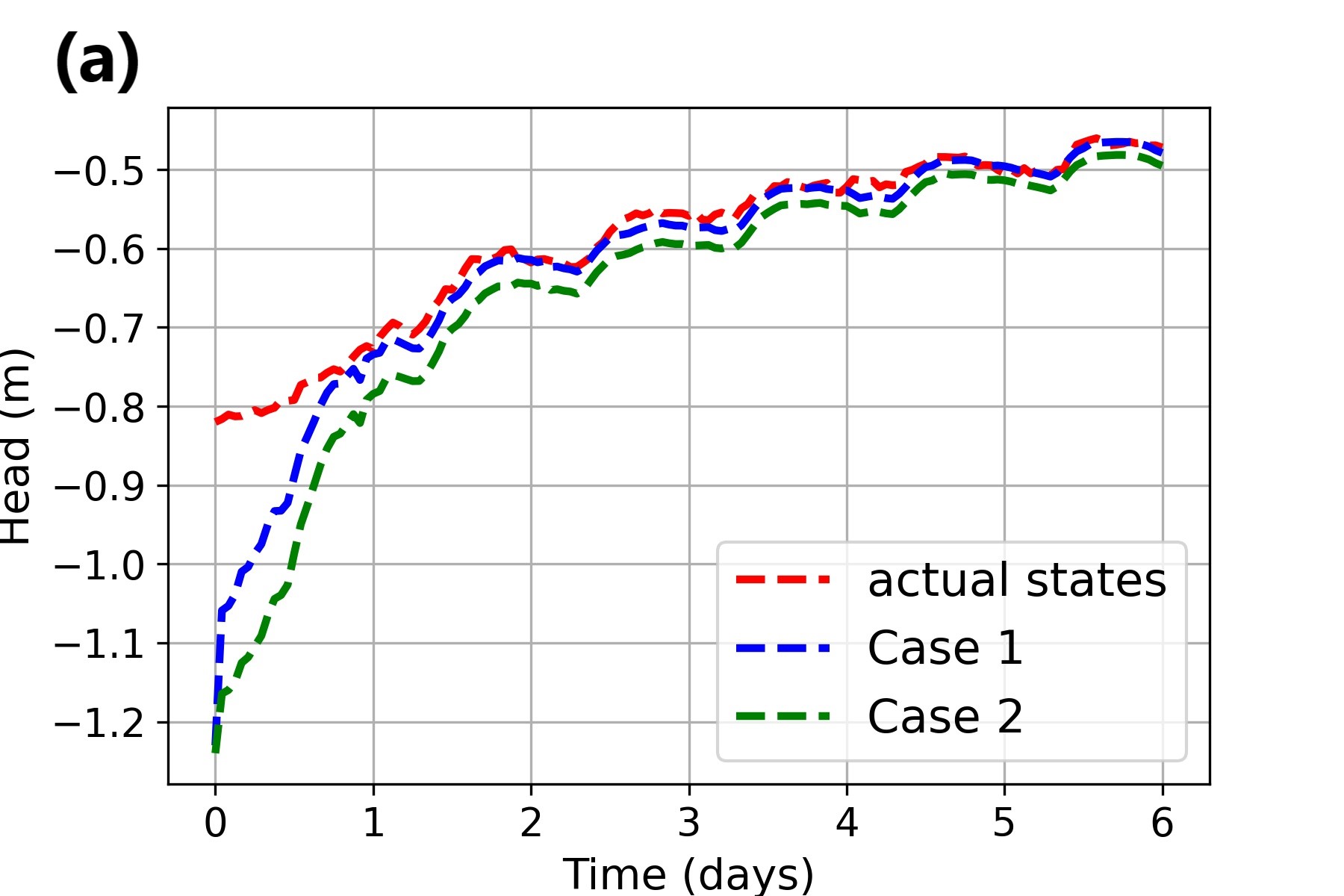}
	}
	\qquad
	\subfloat[State trajectory at depth = 15 cm]{
		\includegraphics[width=0.46\textwidth]{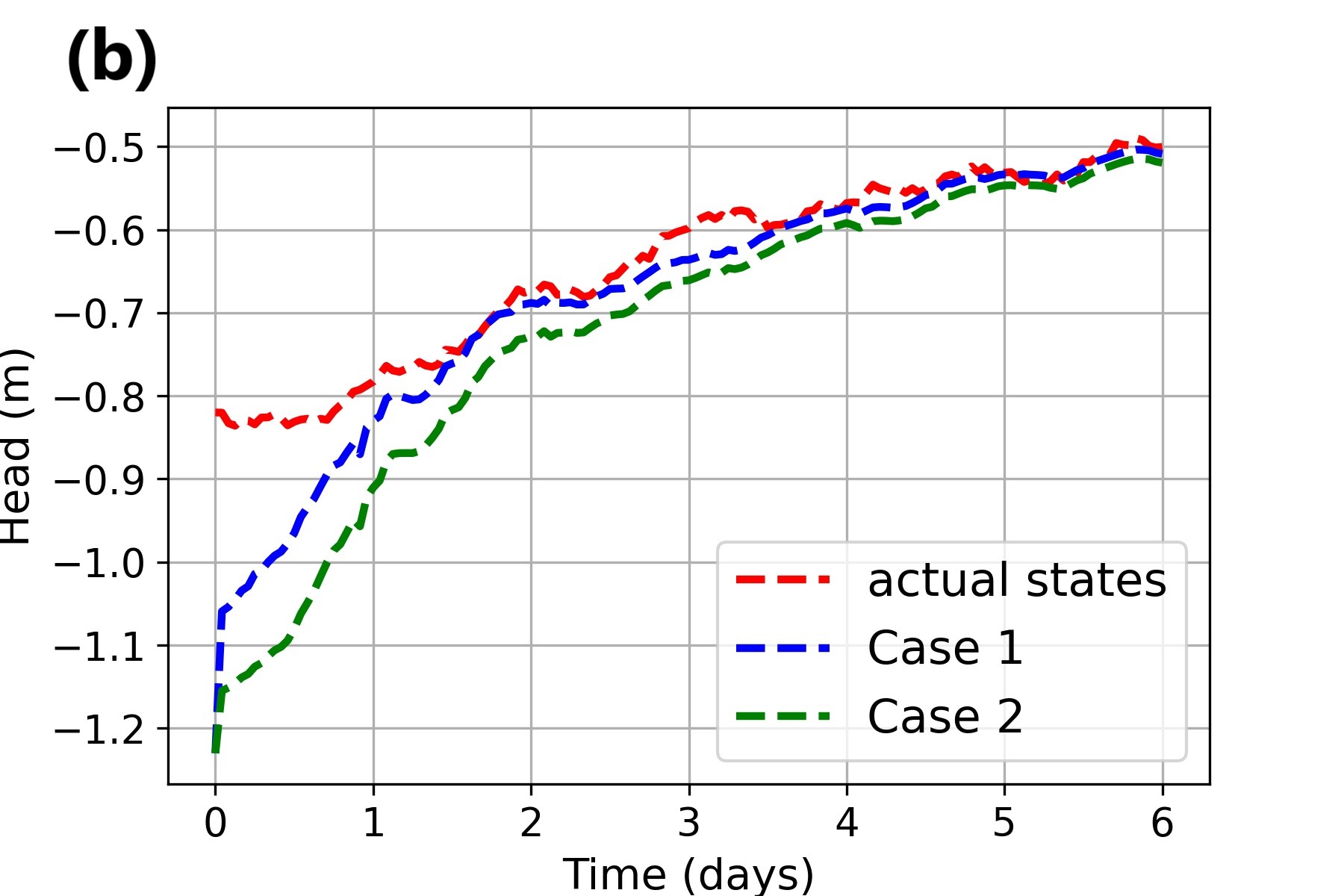}
	}
	
	\centering
	\subfloat[State trajectory at depth = 30 cm]{
		\includegraphics[width=0.46\textwidth]{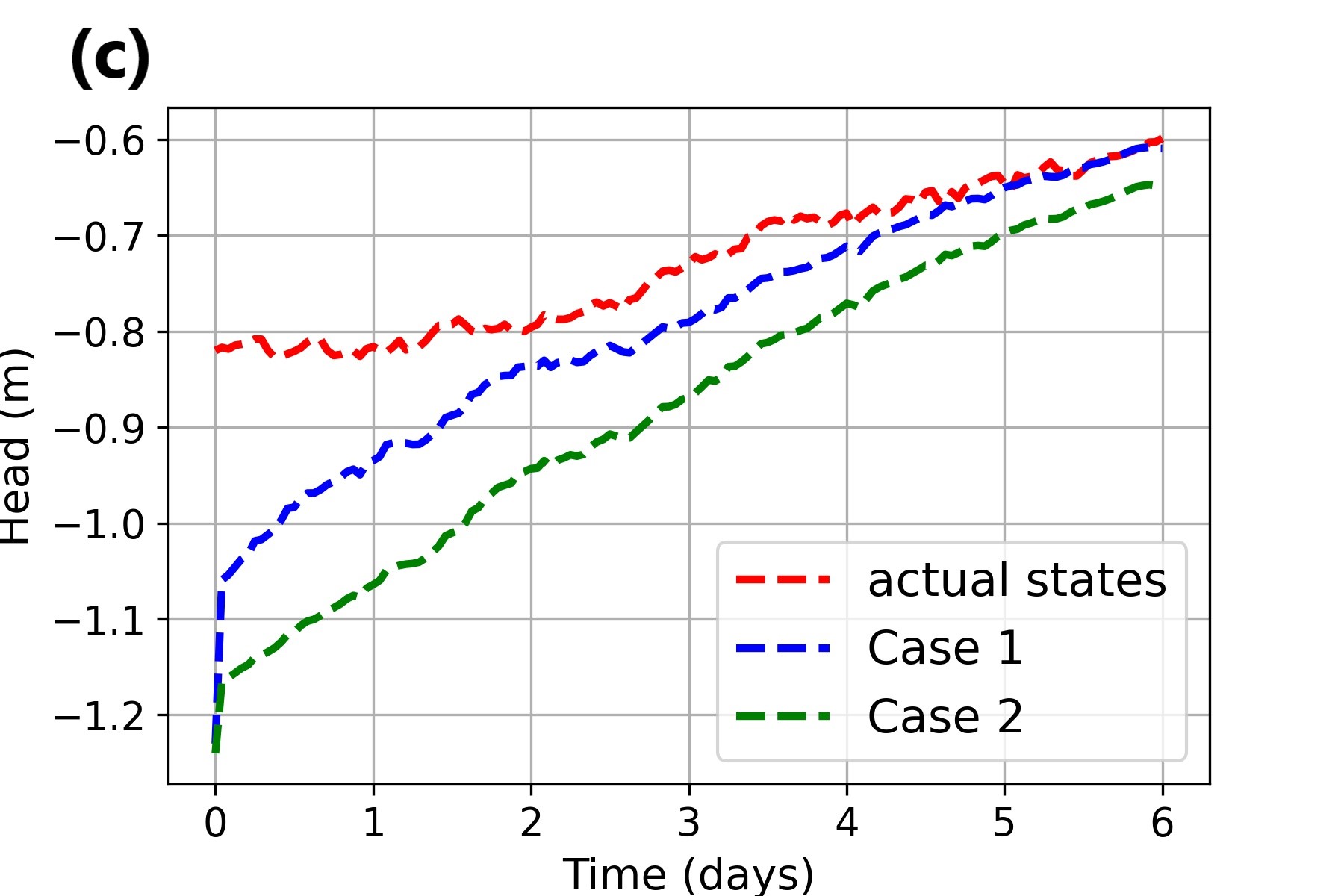}
	} %
	\qquad
	\subfloat[Total estimation error trajectory]{
		\includegraphics[width=0.46\textwidth]{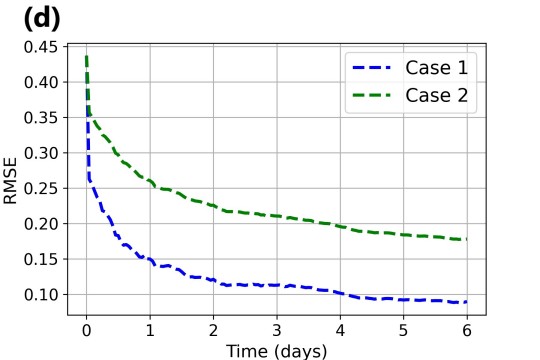}
		
	}
	\caption{Trajectories of actual states and estimated states at some testing nodes}
	\label{fig:result3}
\end{figure}

To verify the effectiveness of the proposed method, two different cases are considered.
In the first case, the sensors are placed at 12 nodes with a higher degree of observability, around 1.8056, and the second case is where 12 sensors correspond to nodes with a lower degree of observability, about 0.2937. 
Figure \ref{fig:result3}, represents the trajectories of the actual states and estimated states for cases 1 and 2 at some testing nodes.
From Figure \ref{fig:result3}, firstly it can be seen that the EKF (blue, green dash-dot) estimates are able to track the actual process states (red dash-dot) very well. Secondly, it can be observed that the estimates by placing the sensors with higher degree of observability converge faster to the actual states. Figure 7(d) compares the total estimation error between case 1 and case 2 and it demonstrates that the root mean square error (RMSE) in case 1 is smaller than case 2 over the simulations.
Also, the average NRMSE over 6 days simulations in case 1 is 15.95\% while in case 2 is 28.70\%. Thus, optimally placed sensors can improve the soil moisture estimation performance for the three-dimensional agro-hydrological system with heterogeneous soil parameters and initial conditions when the simulated data is used.

\begin{figure}[h!]
	\centering
	\subfloat[Actual map]{
		\includegraphics[width=0.28\textwidth]{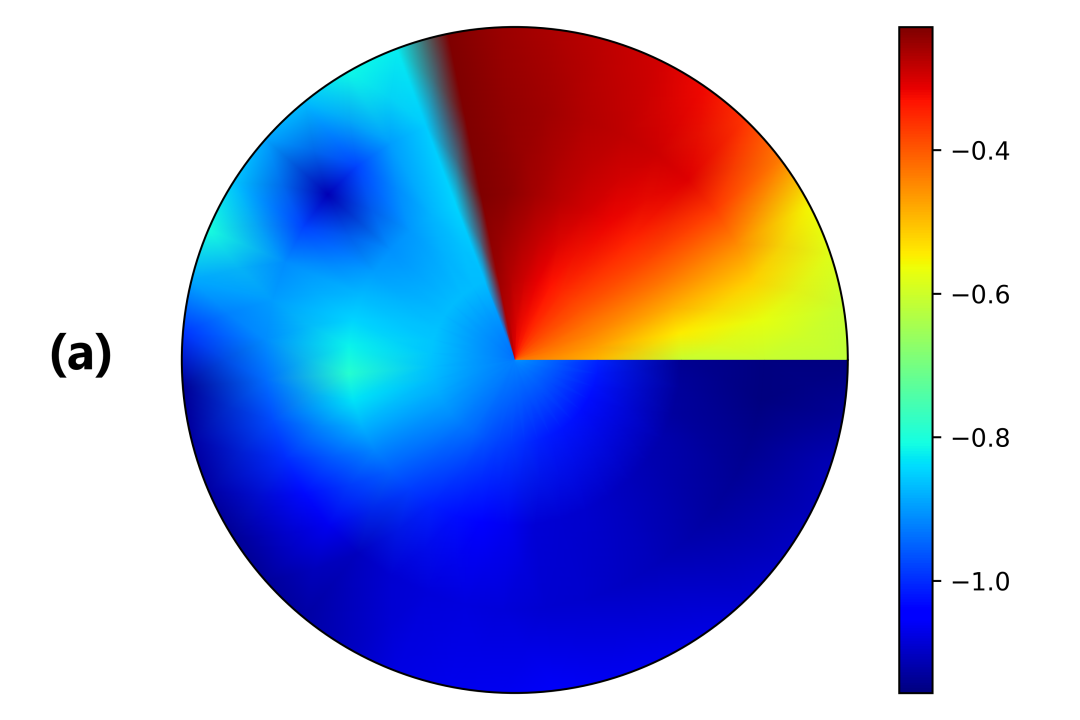}
	} 
	\qquad
	\subfloat[Estimated map]{
		\includegraphics[width=0.28\textwidth]{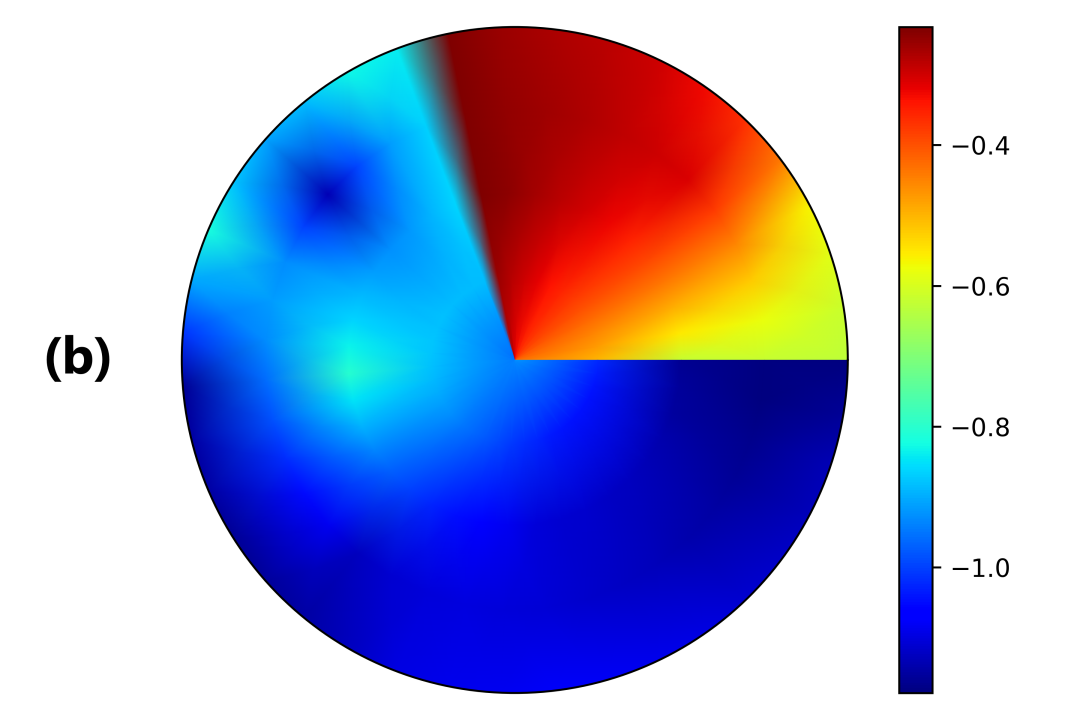}
	}
	\qquad
	\subfloat[Absolute error map]{
		\includegraphics[width=0.28\textwidth]{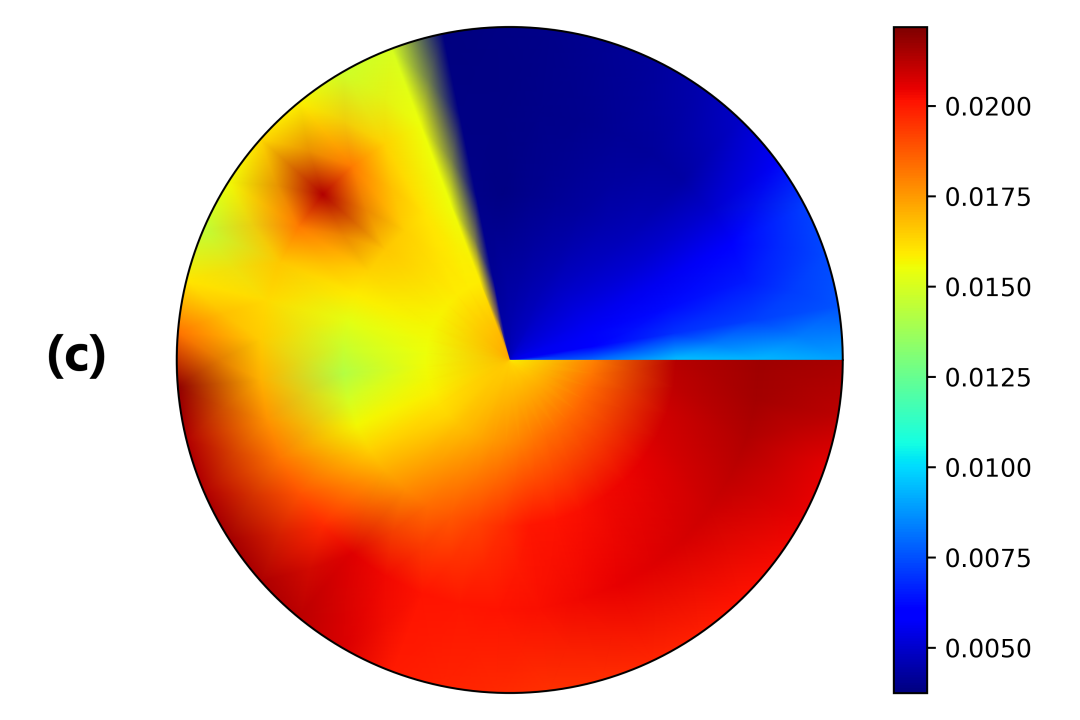}
	}
	\caption{Surface soil water content maps at 02:24 HRS on Day 2 in case 1}
	\label{fig:map1}
\end{figure}

\begin{figure}[h!]
	\centering
	\subfloat[Actual map]{
		\includegraphics[width=0.28\textwidth]{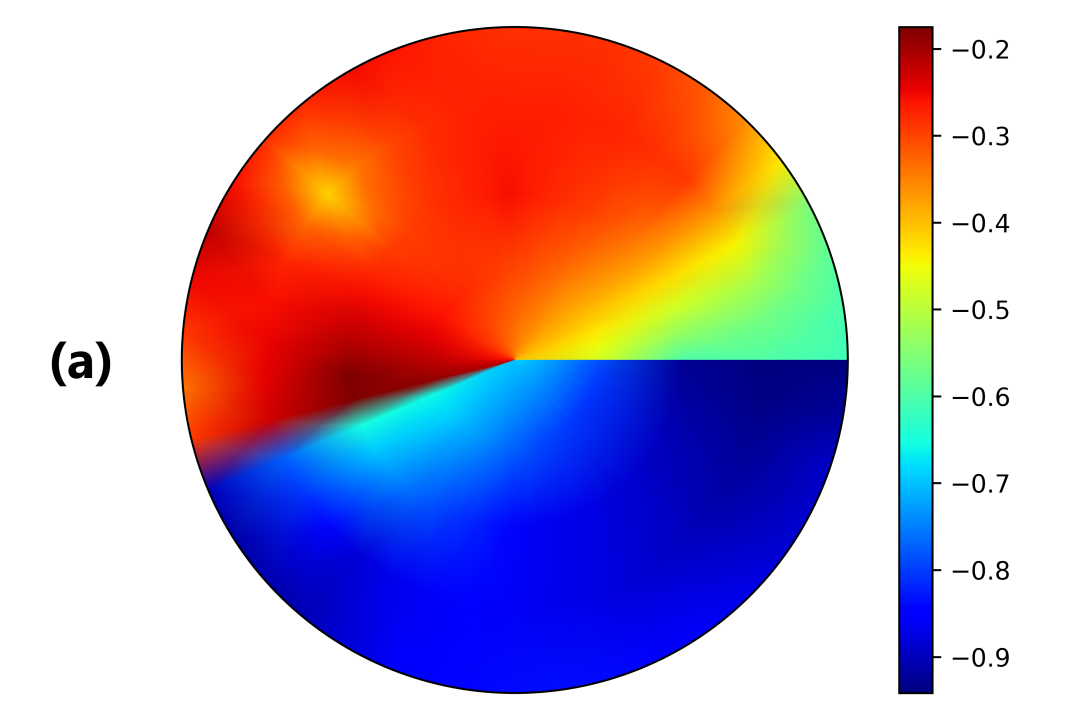}
	} 
	\qquad
	\subfloat[Estimated map]{
		\includegraphics[width=0.28\textwidth]{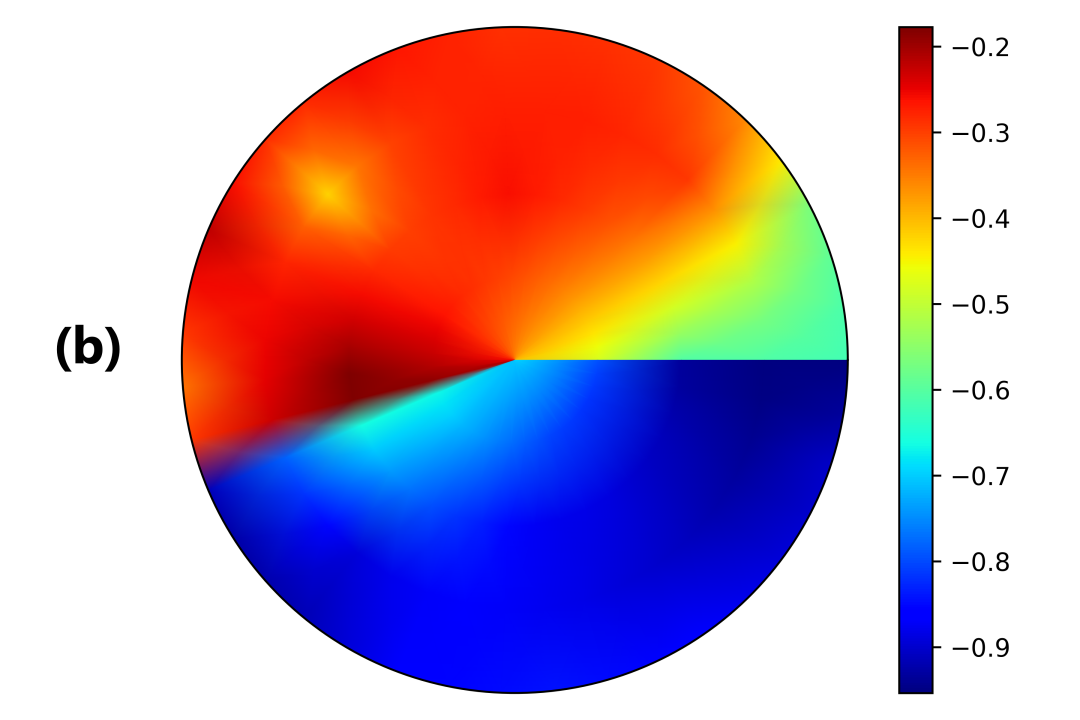}
	}
	\qquad
	\subfloat[Absolute error map]{
		\includegraphics[width=0.28\textwidth]{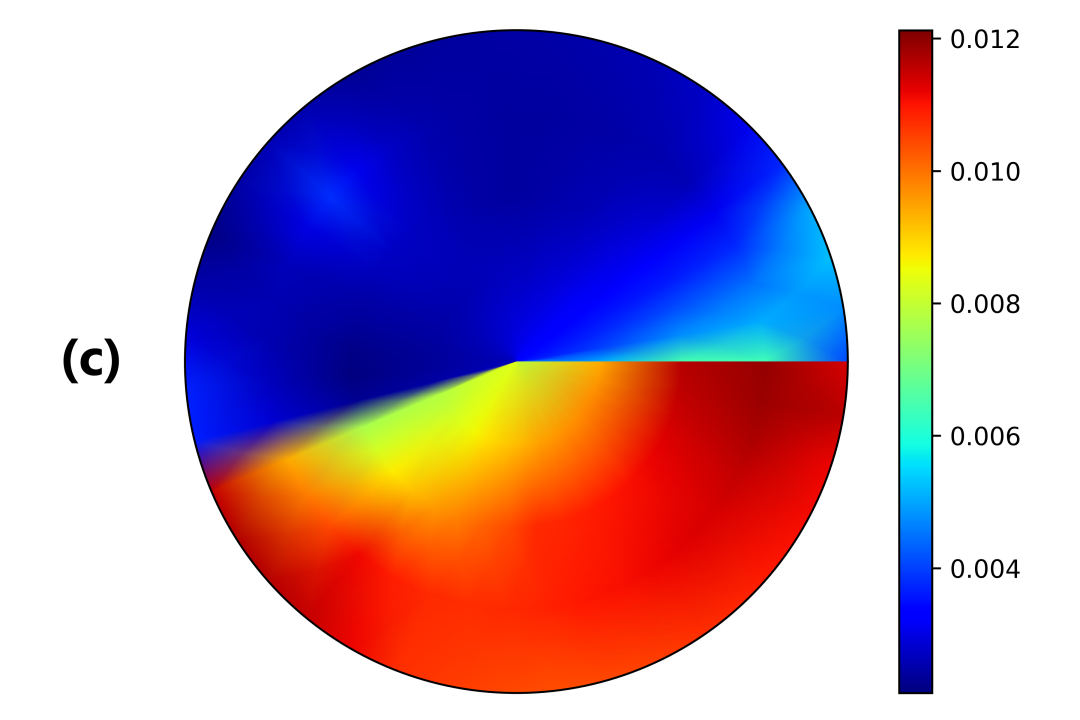}
	}
	\caption{Surface soil water content maps at 04:24 HRS on Day 3 in case 1}
	\label{fig:map2}
\end{figure}

\begin{figure}[h!]
	\centering
	\subfloat[Actual map]{
		\includegraphics[width=0.28\textwidth]{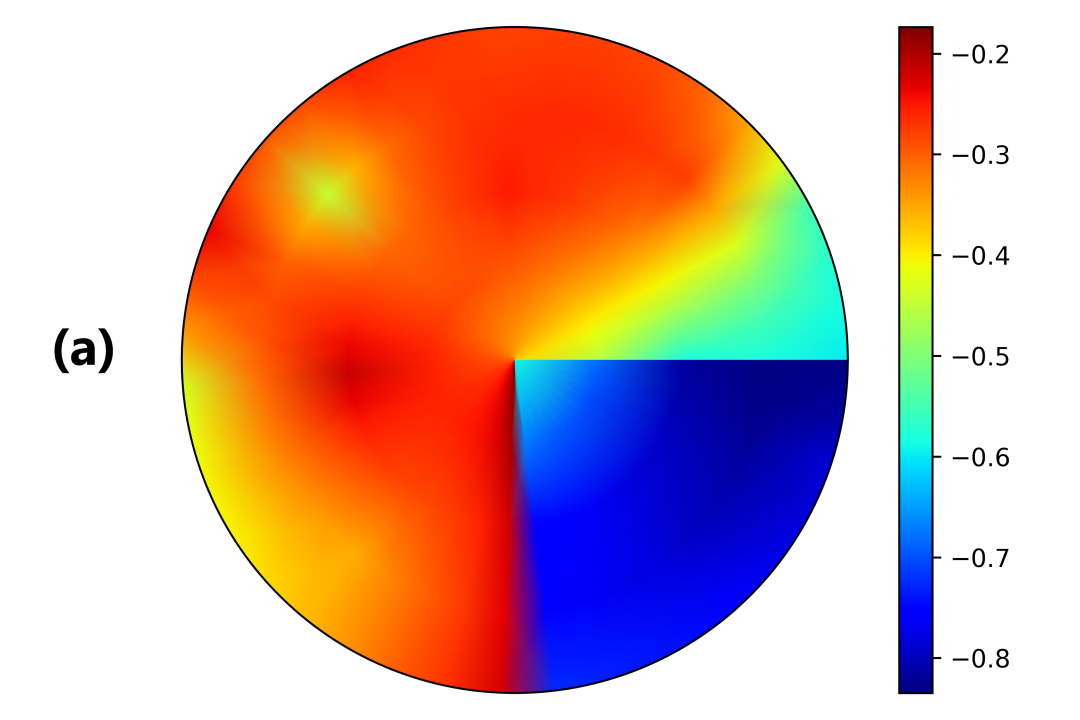}
	} 
	\qquad
	\subfloat[Estimated map]{
		\includegraphics[width=0.28\textwidth]{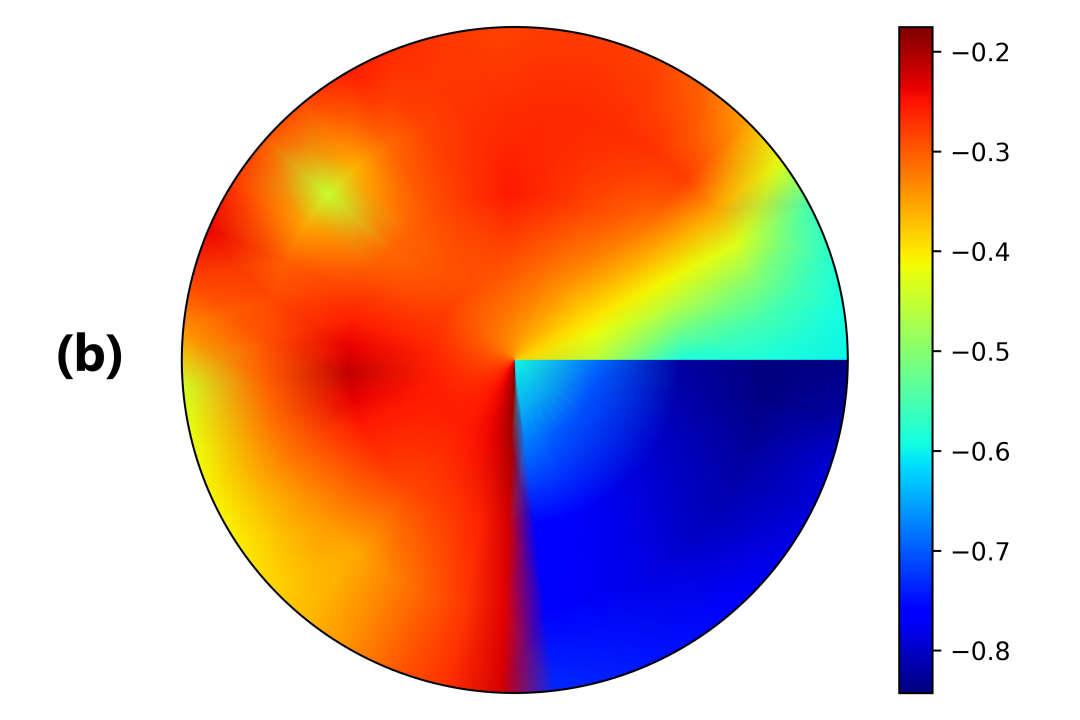}
	}
	\qquad
	\subfloat[Absolute error map]{
		\includegraphics[width=0.28\textwidth]{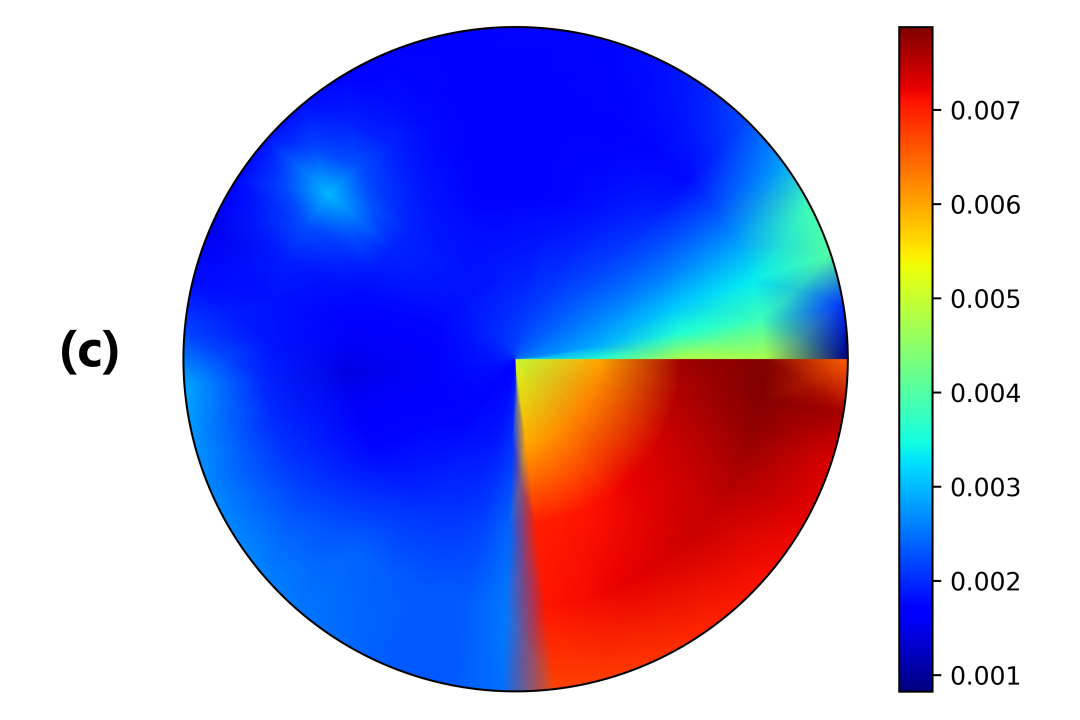}
	}
	\caption{Surface soil water content maps at 06:00 HRS on Day 4 in case 1}
	\label{fig:map3}
\end{figure}

\begin{figure}[h!]
	\centering
	\subfloat[Actual map]{
		\includegraphics[width=0.27\textwidth]{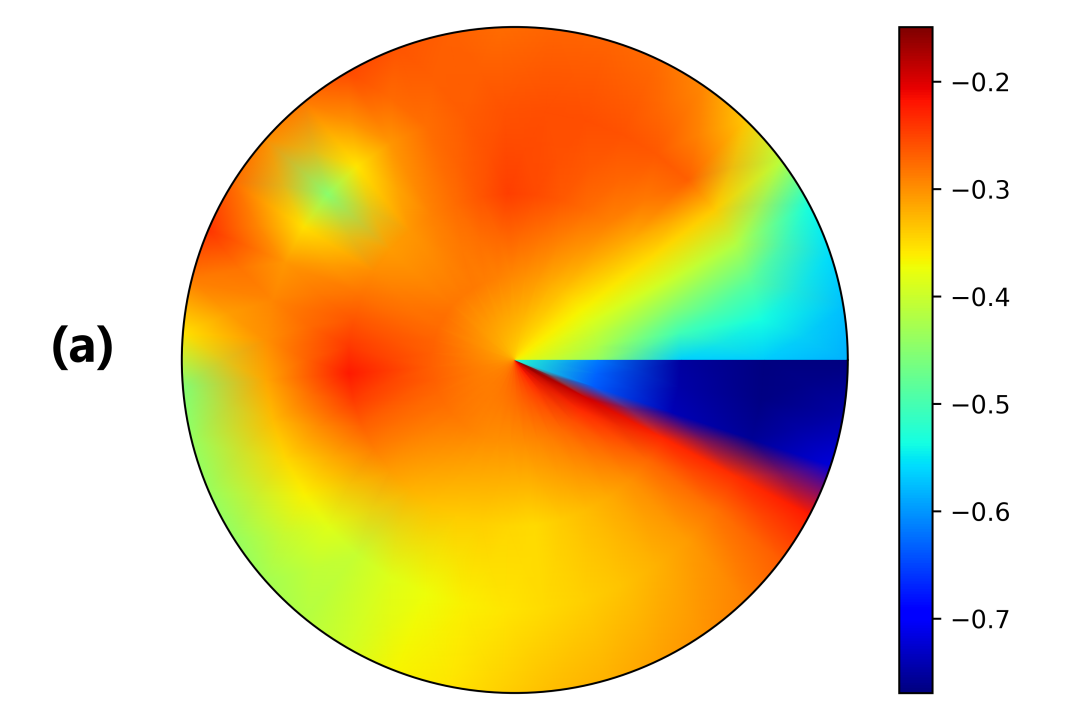}
	} 
	\qquad
	\subfloat[Estimated map]{
		\includegraphics[width=0.27\textwidth]{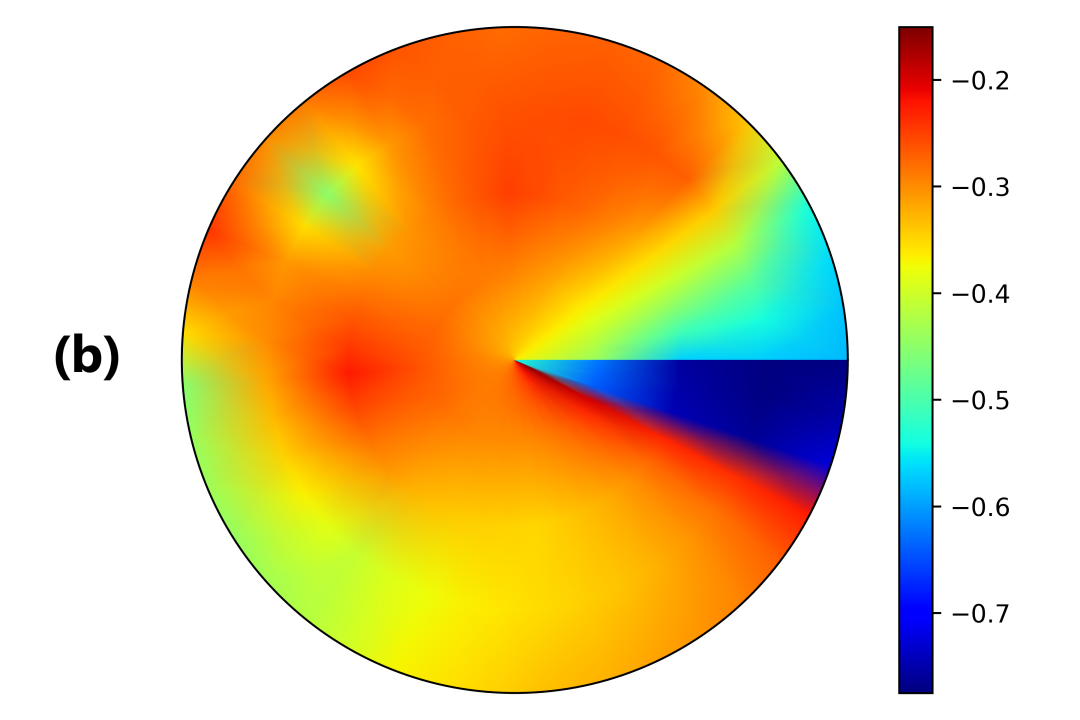}
	}
	\qquad
	\subfloat[Absolute error map]{
		\includegraphics[width=0.27\textwidth]{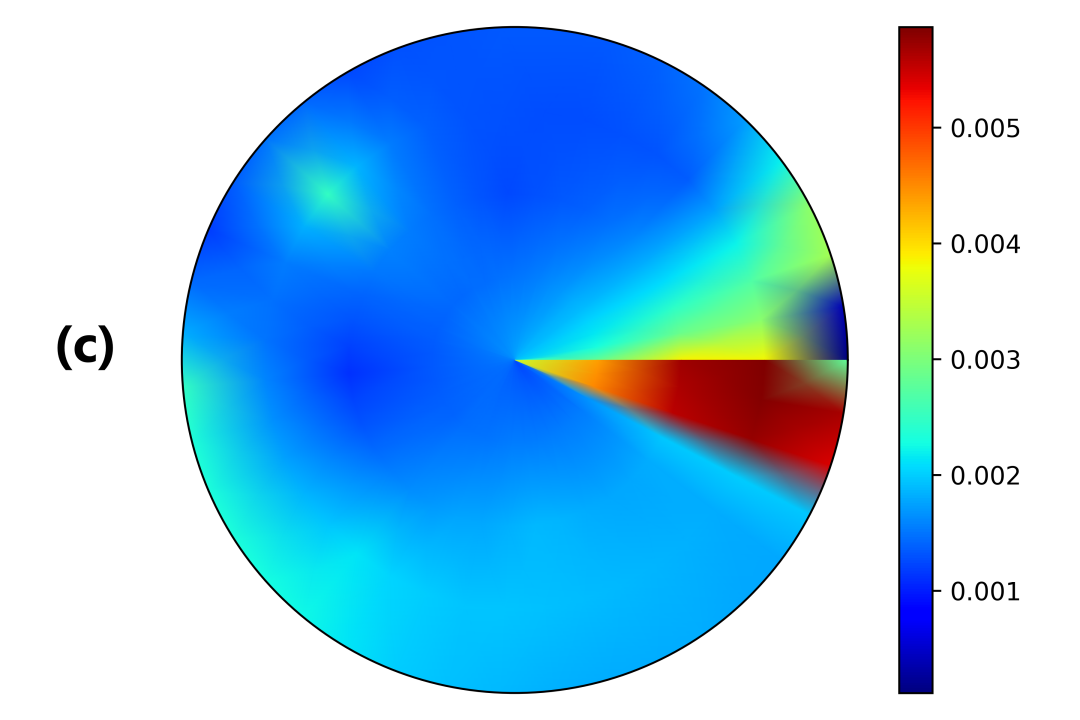}
	}
	\caption{Surface soil water content maps at 07:24 HRS on Day 5 in case 1}
	\label{fig:map4}
\end{figure}

\begin{figure}[h!]
	\centering
	\subfloat[Actual map]{
		\includegraphics[width=0.27\textwidth]{2xx.png}
	} 
	\qquad
	\subfloat[Estimated map]{
		\includegraphics[width=0.27\textwidth]{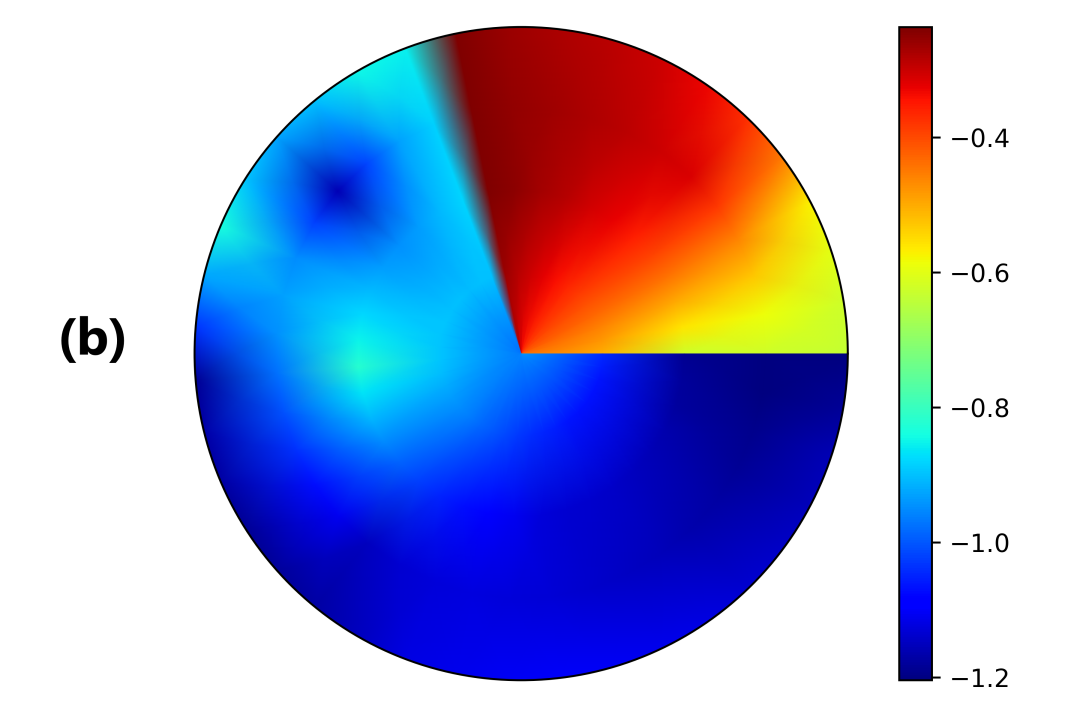}
	}
	\qquad
	\subfloat[Absolute error map]{
		\includegraphics[width=0.27\textwidth]{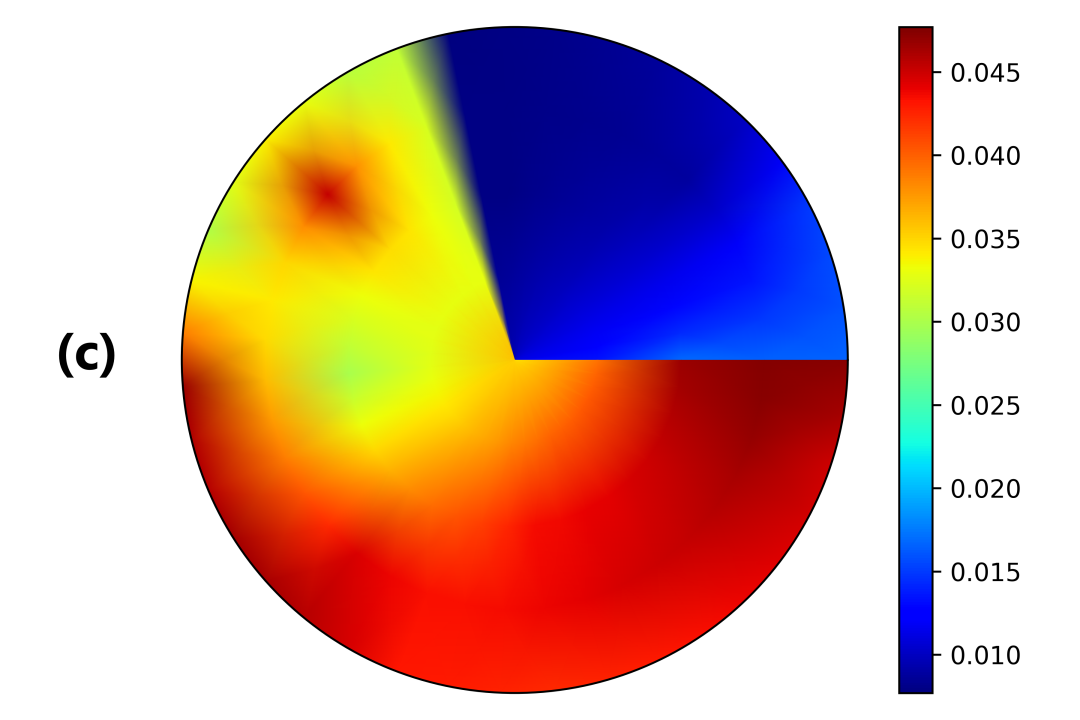}
	}
	\caption{Surface soil water content maps at 02:24 HRS on Day 2 in case 2}
	\label{fig:map5}
\end{figure}

\begin{figure}[h!]
	\centering
	\subfloat[Actual map]{
		\includegraphics[width=0.27\textwidth]{3xx.png}
	} 
	\qquad
	\subfloat[Estimated map]{
		\includegraphics[width=0.27\textwidth]{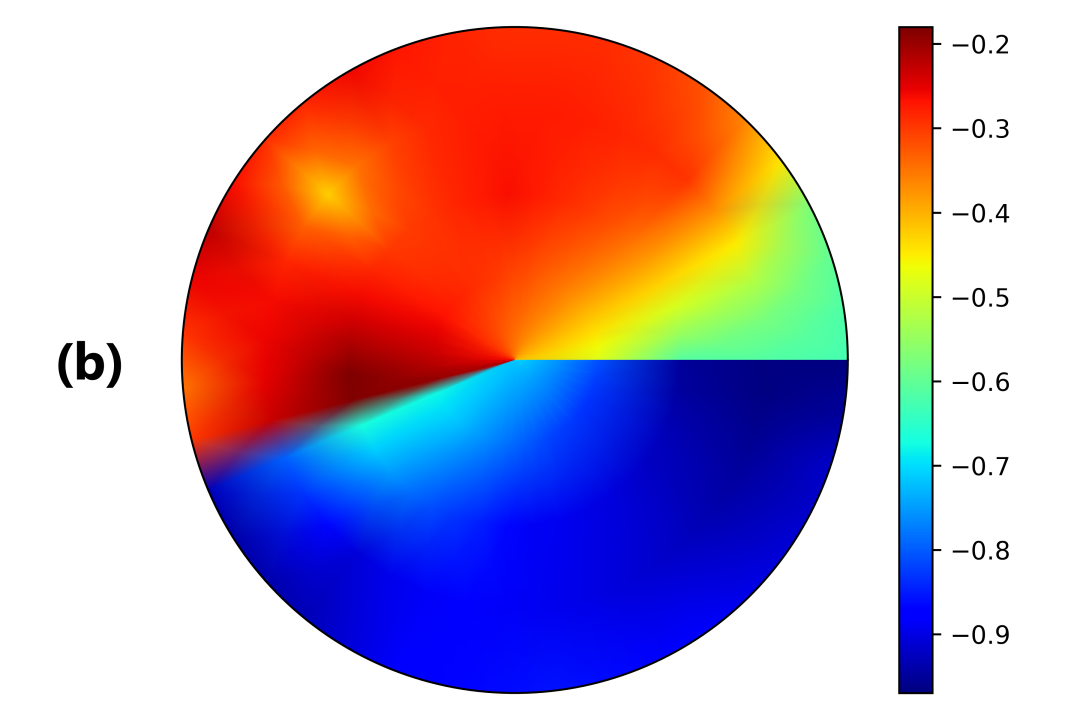}
	}
	\qquad
	\subfloat[Absolute error map]{
		\includegraphics[width=0.27\textwidth]{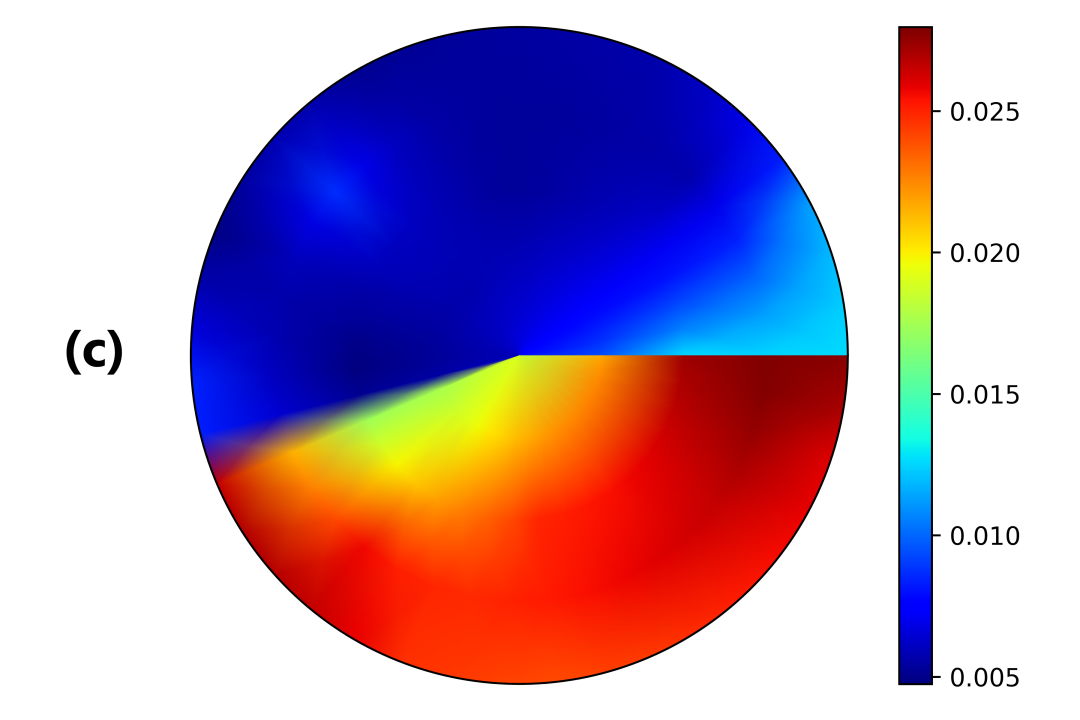}
	}
	\caption{Surface soil water content maps at 04:24 HRS on Day 3 in case 2}
	\label{fig:map6}
\end{figure}

\begin{figure}[h!]
	\centering
	\subfloat[Actual map]{
		\includegraphics[width=0.27\textwidth]{4xx.png}
	} 
	\qquad
	\subfloat[Estimated map]{
		\includegraphics[width=0.27\textwidth]{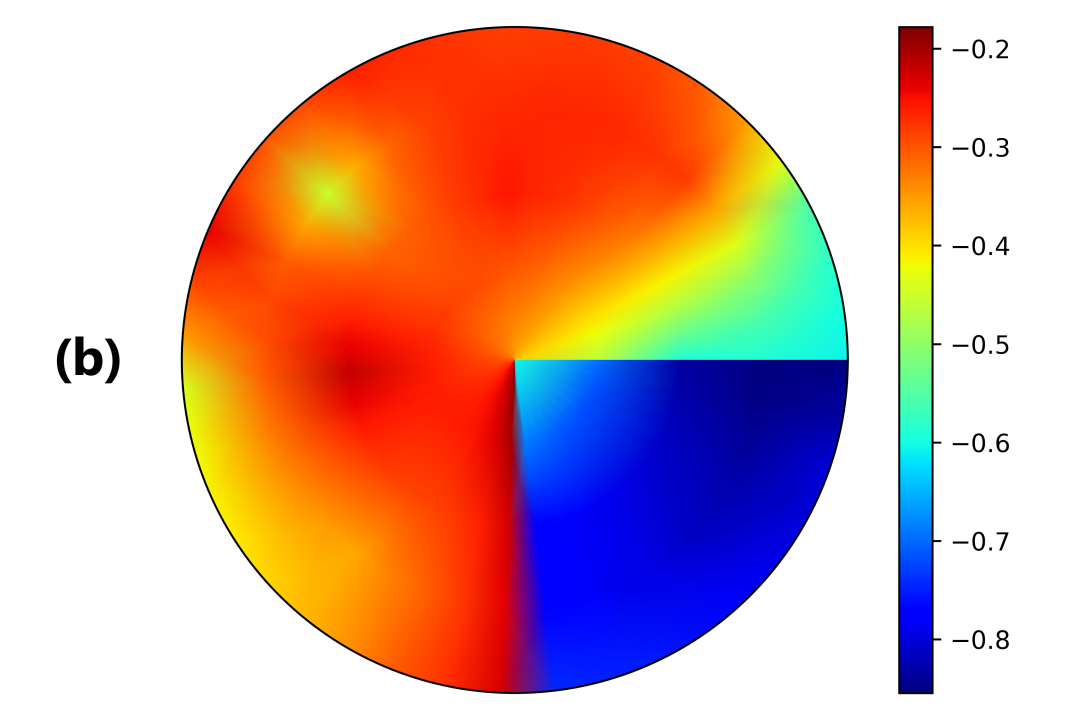}
	}
	\qquad
	\subfloat[Absolute error map]{
		\includegraphics[width=0.27\textwidth]{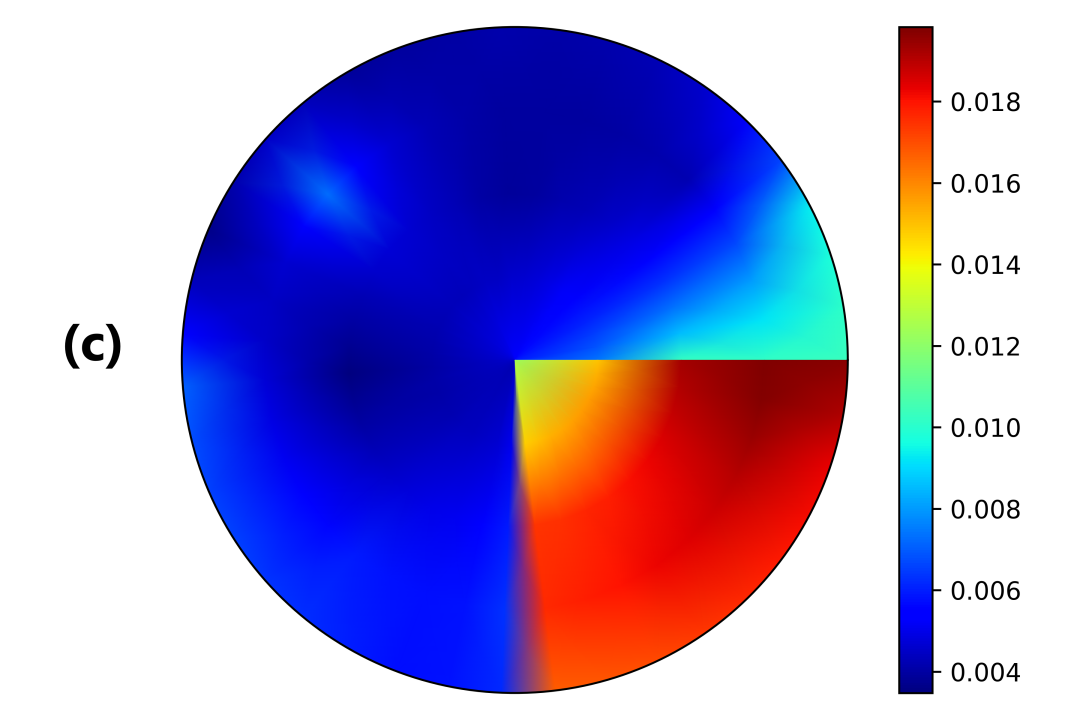}
	}
	\caption{Surface soil water content maps at 06:00 HRS on Day 4 in case 2}
	\label{fig:map7}
\end{figure}

\begin{figure}[h!]
	\centering
	\subfloat[Actual map]{
		\includegraphics[width=0.27\textwidth]{5xx.png}
	} 
	\qquad
	\subfloat[Estimated map]{
		\includegraphics[width=0.27\textwidth]{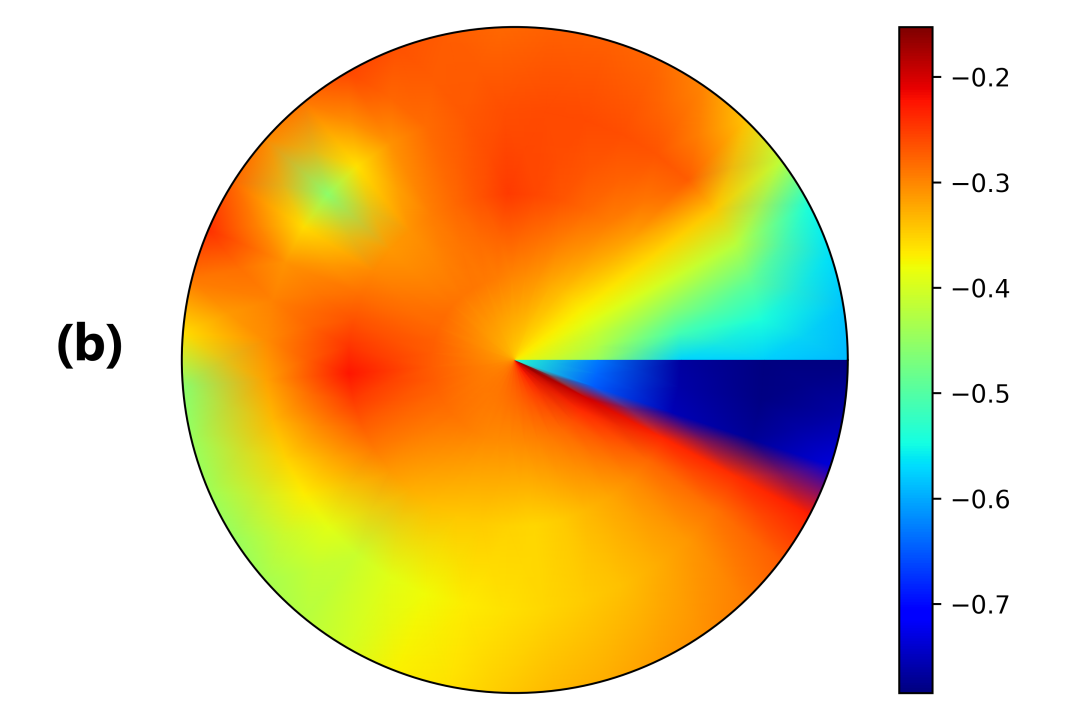}
	}
	\qquad
	\subfloat[Absolute error map]{
		\includegraphics[width=0.27\textwidth]{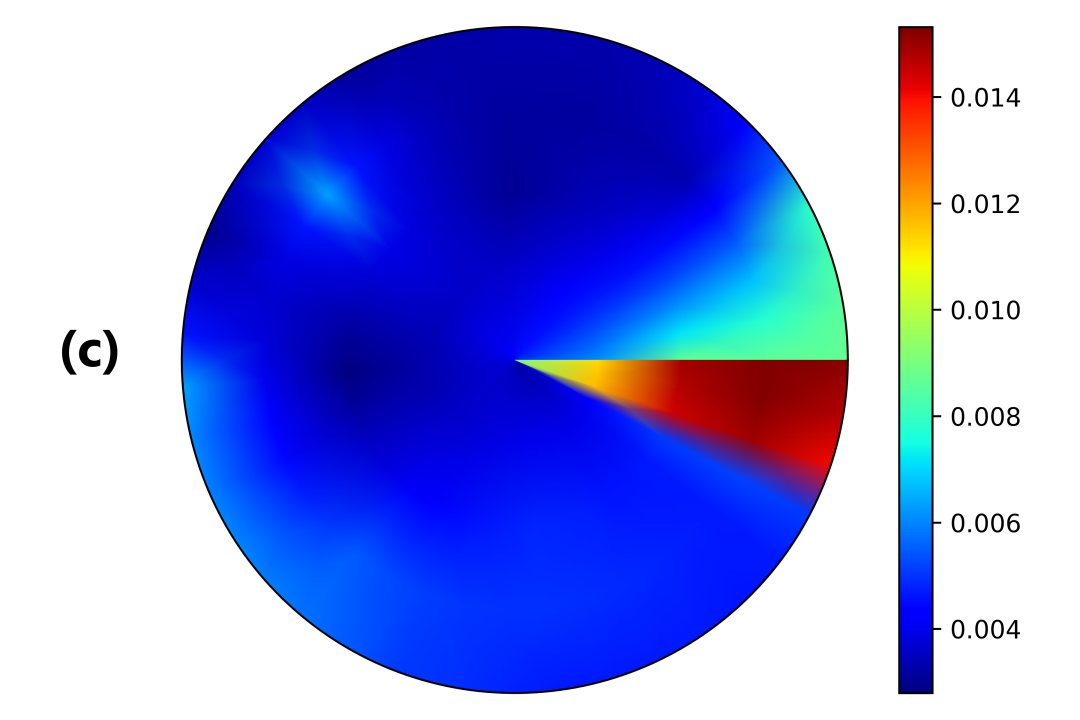}
	}
	\caption{Surface soil water content maps at 07:24 HRS on Day 5 in case 2}
	\label{fig:map8}
\end{figure}

As a further analysis, we construct the actual and estimated soil water content maps to examine the performance of state estimation for a large number of states. We also construct the absolute error maps by computing the absolute error ($e_{k}$) between the actual soil water content and the estimated soil water content.
\begin{equation}
    e_{k} = x_{k} - \hat{x}_{k}
\end{equation}
Figures \ref{fig:map1}-\ref{fig:map4} represent the soil water content maps constructed at selected times during the simulation period for the surface of the field in case 1 where the optimal sensor placement is considered. 
From the Figures, it is observable that the agreement between the estimated maps and the actual maps is significantly strengthened as the simulation time proceeds. Specifically, based on the Figures \ref{fig:map1} and \ref{fig:map4}, the range of the absolute error on the second day is between 0.005 and 0.02, while on the fifth day it decreases to the range of 0.001 and 0.005.   
In addition, Figures \ref{fig:map5}-\ref{fig:map8} indicate the soil water content maps at the same times for the surface of the field in case 2. 
By comparison Figures \ref{fig:map1}-\ref{fig:map4} to Figures \ref{fig:map5}-\ref{fig:map8}, it can be seen that the absolute error maps in case 1 have smaller values compared to the absolute error maps in case 2 at the same times.  
Thus the EKF estimation with optimally sensor placement is able to provide more accurate soil water content maps.


\clearpage

\section{Validation of sensor placement using real data} 
\label{Section 7}

In this section, we investigate the impact of optimal sensor placement in soil water estimation of the studied field using real collected data under different scenarios.
The collected data includes the soil water tension of 14 locations at depth of 25 cm, 14 locations at depth of 50 cm, and 14 locations at depth of 75 cm. 
Before using the collected data as the measurements in the EKF, we have performed some preprocessing steps. First, we converted the soil water tension (Kpa) into the soil head pressure (m). Then, we normalized data using min-max normalization and transformed data between zero and one. It should be pointed out that normalized values are not allowed to be used in the model of the system, Eq. \ref{eq:original system}. Thus in Richards' equation, we use the non-normalized values so that the model realizes how really dry or wet the field is. In addition, we analyzed the data set to determine which areas of the field have been irrigated over the time period of the experiment.
Table \ref{tbl:irrigation} shows the amount and time of irrigation applied to the studied field over the time period.

\begin{table}[H]
	\caption{Irrigation amount and scheduling of the studied field}
	\small 
	\centering
	\begin{tabular}{cccccc}
		\hline
		 {Date} & {July 4} & {July 18} & {July 26} & {July 30} & {August 6}\\
		\hline
		Amount (mm) & 1.81  & 1.58 & 1.58 & 1.51 & 3.16\\
		\hline
	\end{tabular} \label{tbl:irrigation}
\end{table}

Within the simulation period at the sampling time without measurements, the soil moisture predictions are only provided by the field model. When the measurements are available, the head pressure measurements are assimilated into the field model using the EKF. Thus in the presence of the measurements, the soil moisture predictions provided by the field model are updated using the new measurements in the EKF. 
Because of the poor knowledge of the initial state values in the actual application, a wider range of the initial conditions is selected in the real data case. Thus in the following simulation, the initial guess of the state is considered as a random value between -6 m and -5 m. Two scenarios are constructed based on the availability of the number of actual measurements in the studied field.

\subsection{Scenario 1: In presence of two measurements}

In the first scenario, we consider only two sensors in the field. Thus in the update step of EKF, there are only two measurements to correct the prediction values. To observe the effect of sensor placement on the performance of state estimation, we construct two cases. In case 1, the location of the sensor is determined based on the optimal sensor placement result while in case 2, the sensor position is selected randomly. Thus, in the first case, $x_{2}$ and $x_{213}$ with the highest degree of observability among the measurement nodes are selected as the location of the sensors and in the second case, the sensors are placed at $x_{3393}$ and $x_{3542}$ with a lower degree of observability.

\begin{figure}
	\centering
	\subfloat[State trajectory at depth = 25 cm]{ 
	\includegraphics[width=0.45\textwidth]{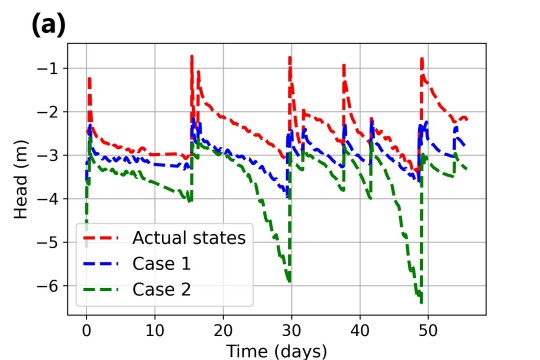}
	}
	\qquad
	\subfloat[State trajectory at depth = 50 cm]{
		\includegraphics[width=0.45\textwidth]{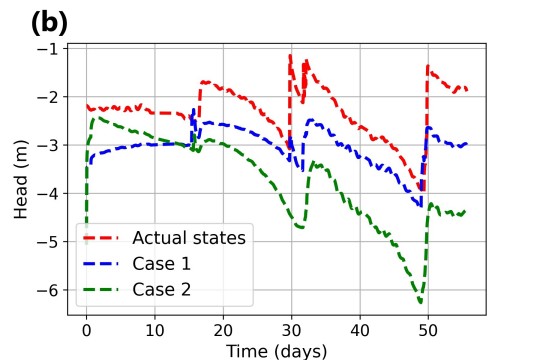}
	}
	
	\centering
	\subfloat[State trajectory at depth = 50 cm]{
		\includegraphics[width=0.45\textwidth]{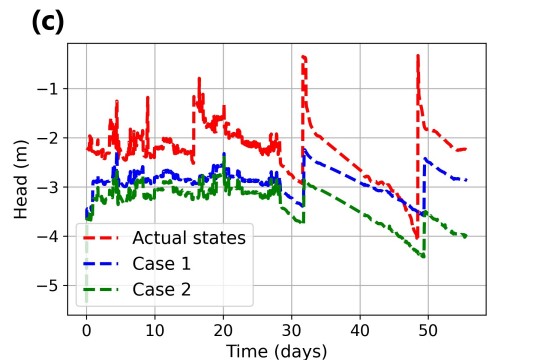}
	} %
	\qquad
	\subfloat[Total estimation error trajectory]{
		\includegraphics[width=0.45\textwidth]{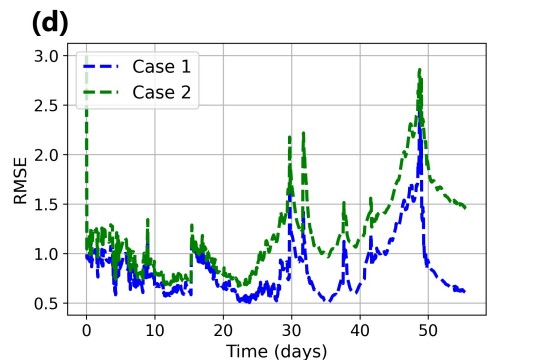}
	}
	\caption{Trajectories of real states and estimated states at some validation points in scenario 1}
	\label{fig:result5}
\end{figure}

Figure \ref{fig:result5} shows the trajectories of the real states (red dash-dot) and estimated states (blue, green dash-dot) at some validation points. 
From Figures \ref{fig:result5}, it can be seen that the estimates by placing the sensors with higher degree of observability (case 1) converge faster to the actual states. Figure 16(d) compares the total estimation error between case 1 and case 2 and it demonstrates the RMSE in case 1 is smaller than case 2 over the simulations.
Also, the average NRMSE over 50 days simulation in case 1, 30.75\%, is much smaller than case 2, 44.56\%.

\subsection{Scenario 2: In presence of a few measurements}
In the second scenario, we consider a few sensors in the studied field. Thus there are more measurements in the EKF to update the soil moisture predictions. 
To verify the effectiveness of the proposed method, two cases are constructed. In case 1, of 42 data points, 15 measured nodes with a higher degree of observability, around 0.8607, are considered as the measurements in EKF. While, in case 2, another 15 measured nodes with a lower degree of observability, about 0.3734, are used as the training points. Also, the rest of the measurements are treated as validation points to compare the real states with estimated states in cases 1 and 2.

\begin{figure}
	\centering
	\subfloat[State trajectory at depth = 25 cm]{ 
	\includegraphics[width=0.45\textwidth]{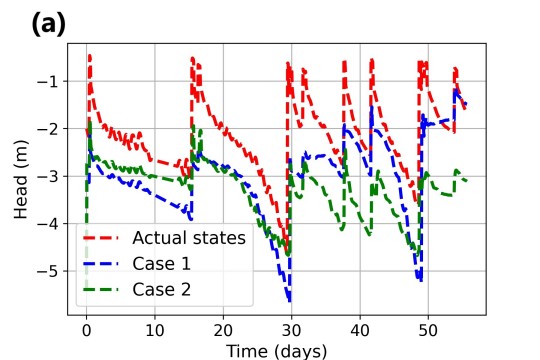}
	}
	\qquad
	\subfloat[State trajectory at depth = 50 cm]{
		\includegraphics[width=0.45\textwidth]{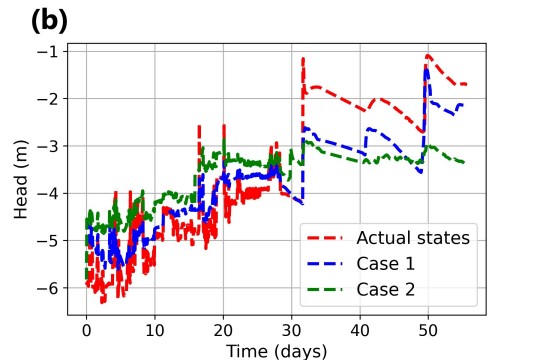}
	}
	
	\centering
	\subfloat[State trajectory at depth = 75 cm]{
		\includegraphics[width=0.45\textwidth]{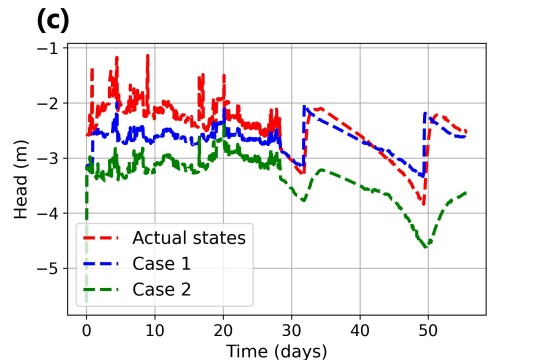}
	} %
	\qquad
	\subfloat[Total estimation error trajectory]{
		\includegraphics[width=0.45\textwidth]{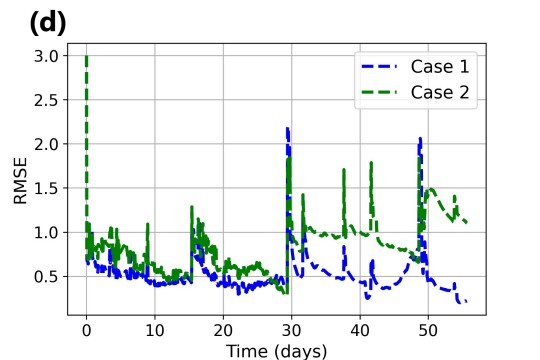}
	}
	\caption{Trajectories of real states and estimated states at some validation points in scenario 2}
	\label{fig:result6}
\end{figure}

Figure \ref{fig:result6}, represents the trajectories of the real states and EKF estimated states at some validation points. From Figures \ref{fig:result6}, firstly it can be seen that, by increasing the number of measurements in the second scenario, the performance of state estimation is improved. In addition, based on Figure \ref{fig:result6}, the performance of the state estimation is significantly improved in case 1 where there are measurements with a higher degree of observability, compared to case 2 that the measurements have a lower degree of observability. 
Figure 17(d) compares the total estimation error between case 1 and case 2 and it demonstrates the RMSE in case 1 is smaller than case 2 over the simulations.
Also, the average NRMSE over 50 days simulation in case 1, 17.11\%, is much smaller than case 2, 27.65\%.
Therefore, state estimation with optimally sensor placement is able to provide more accurate estimates in the actual application.

In the end, we compare the simulation case study and real data case over the same simulation days using the average normalized RMSE. The NRMSE in the simulation study for case 1 (optimally placed sensors) and case 2 (sensors with a lower degree of observability) over 10 days simulation is about 13.69\% and 25.70\% respectively, while NRMSE in the real data study for cases 1 and 2 is respectively 17.36\% and 26.12\%, over the same simulation days.
These comparisons demonstrate that optimal sensor placement can significantly improve the performance of state estimation in the actual application and the amount of improvement in the real data case study is very similar to the simulation study case.

\section{Conclusion}

In this article, the impact of optimal sensor placement in soil water estimation of an actual field was investigated. 
The agricultural field studied in this work was described and information on experiments and collected real data was provided. 
The three-dimensional agro-hydrological system with heterogeneous soils was developed to model the studied field.  
The Kriging interpolation method was implemented to obtain the heterogeneous soil parameters of the studied field. 
The modal degree of observability was applied to the field to determine the optimal sensor placement. 
The EKF was employed to estimate the soil water content of the studied field. 
The results obtained in the simulated case study confirmed that the estimates by placing the sensors with higher degree of observability converge faster to the actual states.
The real case study demonstrated the performance of state estimation with optimally sensor placement is significantly improved in the actual applications. 

\section{Acknowledgements}
Financial support from Natural Sciences and Engineering Research
Council of Canada and Alberta Innovates is gratefully acknowledged

\end{document}